\documentclass[10pt,aps,twocolumn,superscriptaddress,preprintnumbers,nofootinbib]{revtex4-1} 
\usepackage{amsmath,amssymb,graphicx,natbib,bm}
\usepackage{enumerate}
\usepackage{hyperref}
\usepackage{color}

\usepackage{empheq}
\usepackage{slashed}

\newcommand{\beq}{\begin {equation}}  
\newcommand{\eeq}{\end   {equation}} 
\newcommand{\bea}{\begin {eqnarray}} 
\newcommand{\eea}{\end   {eqnarray}}  
\newcommand{\baa}{\begin {array}   } 
\newcommand{\eaa}{\end   {array}   }     
\newcommand{\bit}{\begin {itemize} }
\newcommand{\eit}{\end   {itemize} }
\newcommand{\be }{\begin {equation}} 
\newcommand{\ee }{\end   {equation}}
\newcommand{\nn }{\nonumber        }

\newcommand{\tmbox}[1]{\mbox{\tiny{$#1$}}}

\begin{document}


\preprint{UTTG-19-14, TCC-021-14}

\title{Vector Fermion-Portal Dark Matter: Direct Detection and Galactic Center Gamma-Ray Excess}

\author{Jiang-Hao Yu}
\email{jhyu@utexas.edu}
\affiliation{Theory Group, Department of Physics and Texas Cosmology Center,
\\The University of Texas at Austin,  Austin, TX 78712 U.S.A.}


\begin{abstract}
We investigate a neutral gauge boson $X_\mu$ originated from a hidden $U(1)$ extension of the standard model as the particle dark matter candidate. 
The vector dark matter interacts with the standard model fermions through heavy fermion mediators. 
The interactions give rise to $t$-channel annihilation cross section in the $XX \to f\bar{f}$ process, which dominates the thermal relic abundance during thermal freeze-out and produces measurable gamma-ray flux in the galactic halo.
For a light vector dark matter, if it predominantly couples to the third generation fermions, this model could explain the excess of gamma rays from the galactic center. 
We show that the vector dark matter with a mass of 20 $\sim$ 40 GeV and that annihilate into the $b\bar{b}$ and $\tau\bar{\tau}$ final states provides an excellent description of the observed gamma-ray excess.
The parameter space aimed at explaining the gamma-ray excess, could also provide the correct thermal relic density and is compatible with the constraints from electroweak precision data, Higgs invisible decay, and collider searches. 
We show the dark matter couplings to the nucleon from the fermion portal interactions are loop-suppressed, and only contribute to the spin-dependent cross section.
Therefore the vector dark matter could easily escape the stringent constraints from the direct detection experiments.
\end{abstract}

\maketitle


\section{Introduction}
\label{sec:intro}

The dark matter (DM) provides a natural explanation for approximatively $27\%$ of the composition of the universe. 
Popular particle physics candidate of the DM is the weakly interacting massive particle (WIMP),
which has the annihilation cross section of the order of the electroweak scale to account for the observed relic abundance.

Although various observations of the DM gravitational effects indicate the existence of DM, the search for non-gravitational signals remains one of the most challenging tasks.
Over the decades, there are tremendous progresses in the underground direct detections, in the indirect cosmic ray signals, and at colliders.
The direct detection experiments, through DM scattering off nuclei targets, achieve unprecedented sensitivity to WIMP in the mass range from few GeV to several TeV.
The absence of direct detection signals so far puts strong constraint on the DM coupling to the nucleon. 
In the indirect detection experiments, the DM annihilations or decays in the galactic halo are expected to produce
potentially observable fluxes of high energy particles, including gamma-rays, cosmic rays and also neutrinos.

Of particular interest are gamma rays from the region of the Galactic Center, because the region is predicted to contain very high densities of DM.
Over the past few years~\cite{Goodenough:2009gk,Hooper:2010mq,Hooper:2011ti,Abazajian:2012pn,Hooper:2012sr,Hooper:2013rwa,Gordon:2013vta,Huang:2013pda,Abazajian:2014fta,Daylan:2014rsa}, an excess of gamma rays above the modeled astrophysical emission in the inner region of our galaxy
has been identified in the data of the Fermi Gamma-Ray Space Telescope.
The excess exhibits spherically symmetric spatial distribution, and is uncorrelated with the galactic disk or Fermi bubbles. 
The gamma-ray energy spectrum and the overall rate 
are best fitted by a 30-40 GeV DM annihilating mostly into bottom quarks or 10 GeV DM annihilating significantly into tau leptons~\cite{Daylan:2014rsa}.

From the theoretical point of view, the existence of particle DM indicates a dark sector beyond the standard model (SM) of particle physics. 
The dark sector interacts with the SM particles through mediator particles. 
Since very little is known about its matter content and its interaction, simplified models of DM are quite useful to extract out main features of the underlying dynamics of DM interactions. 
The dark sector and the mediator particles have the following general interactions~\cite{Berlin:2014tja}:
\bea
	{\mathcal{L}}_{s-\textrm{chan}} &=& (\overline{\textrm{DM}}\ \textrm{DM}\ \textrm{mediator}) + (\overline{\textrm{SM}}\ \textrm{SM}\ \textrm{mediator}), \nn\\
	{\mathcal{L}}_{t-\textrm{chan}} &=& (\overline{\textrm{DM}}\ \textrm{SM}\ \textrm{mediator}) + (\overline{\textrm{SM}}\ \textrm{DM}\ \textrm{mediator}).\nn
\eea
The above Lagrangians induce the DM annihilation processes mediated by the $s$-channel or $t$-channel particles. 
A typical $s$-channel mediator is considered to be the Higgs boson, known as the Higgs portal dark matter~\cite{Silveira:1985rk,McDonald:1993ex,Burgess:2000yq,Kim:2006af,Kim:2008pp,Cao:2009uw,Kanemura:2010sh,Mambrini:2011ik}.
As is well-known~\cite{Kanemura:2010sh,Mambrini:2011ik}, the Higgs portal DM 
is highly in tension with the current  direct detection experiments, except for very limited parameter regions: the DM mass is very close to half of the Higgs boson mass, 
or the DM is very heavy.
Recent years, the DM models with $t$-channel mediators becomes the trends~\cite{Agrawal:2011ze,Kile:2011mn,Batell:2011tc,Kamenik:2011nb,Chang:2013oia,Chang:2014tea,Bai:2013iqa,Bai:2014osa,An:2013xka,DiFranzo:2013vra,Papucci:2014iwa} to explain the particle nature of the DM, and to predict distinctive signals in direct, indirect and collider searches. 
Examples are flavored dark matter~\cite{Agrawal:2011ze,Kile:2011mn,Batell:2011tc,Kamenik:2011nb}, effective WIMPs~\cite{Chang:2013oia,Chang:2014tea}, fermion-portal dark matter~\cite{Bai:2013iqa,Bai:2014osa}.
In these models, DM particles are considered to be scalar or fermion particles, and correspondingly, the mediators are fermion or scalar.
The DM particle interacts with the SM fermions through Yukawa coupling to the mediators.
To escape tight constraints from the direct detection experiments, the DM particle either only couples to leptons~\cite{Chang:2014tea,Bai:2014osa}, or the third-generation quarks~\cite{Kumar:2013hfa,Batell:2013zwa}. 
To produce measurable cosmic rays and gamma rays, the DM particle needs to be complex scalar or Dirac fermion.
Among many theoretical explainations~\cite{Berlin:2014tja,Alves:2014yha,Abdullah:2014lla,Martin:2014sxa,Izaguirre:2014vva,Ipek:2014gua,Arina:2014yna,Agrawal:2014una} on the gamma-ray excess, recently a bottom-flavored dark matter model was proposed~\cite{Agrawal:2014una}, although the large coupling strength which is greater than 2 is required.  
To avoid the tension between such large coupling and the stringent direct detection constraints, cancellation among different contributions in the DM-nucleon cross section is also needed in the bottom-flavored model.  
In the bottom-flavored model, the dark matter extensively annihilates into the $b\bar{b}$ final states.
However, the dark matter annihilating into $b\bar{b}$ final state only is in tension with the constraints from the antiproton flux in the indirect detection~\cite{Cirelli:2014lwa,Bringmann:2014lpa}.

We propose a fermion-portal dark matter model, in which the dark matter is a spin-1 vector boson, and the $t$-channel mediators are heavy fermions. 
Vector dark matter has been investigated in many dark matter scenarios, such as the Kaluza-Klein (KK) dark matter~\cite{Servant:2002aq,Cheng:2002ej}, the Little Higgs dark matter~\cite{Birkedal:2006fz,Asano:2006nr},  the vector Higgs portal dark matter~\cite{Hambye:2008bq,Lebedev:2011iq,Farzan:2012hh,Carone:2013wla,Baek:2012se}, and so on.
In the vector Higgs portal model, as discussed above, it is very difficult to both satisfy the direct detection constraints and explain the gamma-ray excess, except that half of the scalar boson mass happens to be the vector dark matter mass within a narrow range~\cite{Ko:2014gha}.
In the universal extra dimension model~\cite{Appelquist:2000nn}, although the KK fermions could serve as the $t$-channel mediators, the KK DM is tightly constrained by the direct detection experiments and thus the DM cannot be light~\cite{Servant:2002hb,Arrenberg:2008wy}.
Similarly in the Little Higgs model with T-parity~\cite{Cheng:2003ju,Cheng:2004yc}, the mass of the T-odd photon DM is also highly constrained and cannot be lighter than 100 GeV~\cite{Wang:2013yba,Chen:2014wua}.
However, to fit the gamma-ray spectrum observed in the galactic center, a light DM with mass around $5-40$ GeV annihilating into the $b\bar{b}$ or the $\tau\bar{\tau}$ final states is needed.
Therefore, we consider a simplified model of the vector dark matter to explain the gamma-ray excess. 
In this simplified model, the vector dark matter interacts with the SM fermions through fermion mediators. 
To explain the gamma-ray excess and evade the direct detection constraints at the same time, we need the DM couples to the third-generation SM fermions extensively. 
This can be satisfied by assuming that the mediators coupled to the first two generation SM fermions are very heavy, and thus decouple.
In this setup, the direct detection bounds are easily evaded, because the coupling of the vector DM to the nucleon is loop-suppressed, and only contributes to the spin-dependent cross section in the DM-nucleon scattering. 
For the vector DM with mass around $5-40$ GeV, the dark matter extensively annihilates into the bottom quarks and the tau leptons. 
Hence, we expect the gamma-ray emission from these final states could explain both the spectrum and the total rate of the gamma-ray excess well.

The organization of the paper is as follows. 
In section II, we outline the vector fermion-portal dark matter model. In Section III, we calculate the thermal relic abundance. In Section IV, both the spin-indepedent and spin-dependent direct-detection cross sections are discussed.  In section V and section VI, we present the constraints from the precision electroweak data, the Higgs invisible decay, and the LHC searches on the mediators.  In section VII, we discuss how the model explain the gamma-ray excess in the galactic center. In Appendix A, a concrete model is presented. In Appendix B, the thermally averaged cross sections are calculated. In Appendix C, we list the effective operators of the dark matter nucleon scattering in the direct detection calculation. In Appendix D, we summarize our calculations on the triangle and box loop induced vertices.


\section{The Model}
\label{sec:model}

The vector dark matter usually comes from spontaneous symmetry breaking of a hidden gauge symmetry.
Let us consider the simplest $U(1)$ extension of the SM gauge group.
The gauge symmetry is $SU(2)_L \times U(1)_Y \times U(1)_{\rm dm}$, 
where the local $U(1)_{\rm dm}$ corresponds to a new gauge boson $X_\mu$.
We introduce a discrete symmetry $Z_2$ to stabilize this new gauge boson as a dark matter candidate.
Under the $Z_2$ symmetry, the new gauge boson has $X_\mu \to - X_\mu$ while all the SM particles  are $Z_2$-even.
After spontaneous symmetry breaking, $X_\mu$ becomes massive through eating the Goldstone component of a complex scalar singlet $\phi$.  
The complex scalar $\phi$ is a SM singlet, and is only charged under under the $U(1)_{\rm dm}$ group. 
To make sure the $Z_2$ symmetry is still exact after electroweak symmetry breaking,  the complex scalar transforms as $\phi \to \phi^*$ under the $Z_2$ symmetry. 
This transformation makes the Goldstone component $Z_2$-odd: $\textrm{Im}\,\phi \to -\textrm{Im}\,\phi$.  
The relevant Lagrangain is
\bea
	{\mathcal L}_{\rm scalar}  = ( \partial^\mu \phi^* - i  g_\phi   X^\mu   \phi^*) (\partial^\mu \phi + i g_\phi   X^\mu   \phi),
\eea
where $g_\phi$ is the coupling of the scalar boson to the vector DM.
After spontaneous symmetry breaking, the gauge boson $X^\mu$ obtains its mass
\bea
	m_{X} =  g_\phi  u,
\eea
where $u$ is the vacuum expectation value corresponding to the $U(1)_{\rm dm}$ symmetry breaking.
Similar to the SM, the dangerous $(\partial^\mu \phi^*) \phi X_\mu$ terms which make the vector DM instable,
disappear after the absorption of the Goldstone boson $\textrm{Im}\,\phi$ by the vector DM. 
Therefore, all the couplings involved in the $X_\mu$ field is in pairs, and thus $X_\mu$ is a stable dark matter candidate.

After symmetry breaking,  the neutral component of the doublet ${\textrm Re}\,H_0$ and the real component of the scalar ${\textrm Re}\,\phi$ mix into the mass eigenstates $(h, S)$ with mixing angle $\varphi$. 
Due to the Higgs-scalar mixing, the vector dark matter couples to both the Higgs boson $h$ and the new scalar $S$:  
\bea
	{\mathcal L}_{XXS} &=& \frac{g_\phi^2}{2} X_\mu X^\mu (- s_\varphi h + c_\varphi S + u)^2, 
\eea
where we denote $s_\varphi \equiv \sin \varphi$ and $c_\varphi \equiv \cos \varphi$.

In this study, we consider the vector dark matter couples to the right-handed SM fermions through the fermion mediators. 
To preserve gauge symmetries, the mediators must have the same quantum numbers as the SM fields. 
In our setup, we take the fermion mediator $F$ as the vector-like singlet fermion, which is considered to be the partner of the SM fermion $f$. 
We assume that the interaction is invariant under the discrete $Z_2$ symmetry. 
This implies that similar to the vector dark matter, the fermion singlet $F$ is $Z_2$-odd under the discret symmetry. 
Following the framework of the simplified models of DM,
the couplings of the SM fermion and its partner to the vector dark matter can be written as
\bea
	\mathcal{L}_{\rm fermion} &=&  \overline{F_a} i\gamma_\mu \left[\partial^\mu + i e Q_a (A^\mu - \tan\theta_W Z^\mu) \right]  {F_a} \nn\\
	&& -  g_{Xa} \overline{F}_a \gamma_\mu   P_R {f}_a X^\mu + h.c.,
\eea
where $Q$ is the electric charge of the mediator $F_a$ with $a$ the flavor index, $g_{Xa}$ is the new gauge coupling,  and $P_R = \frac{1+\gamma_5}{2}$ is the right-handed chiral projector.
A concrete model on how to construct the coupling is described in Appendix A. 
Here we neglect the mixing among the three SM generations, and thus the coupling $g_X$ is flavor diagonal in our setup. 
Regarding to the spectrum in the dark sector,  
the DM particle is completely stable as long as it is lighter than mediator,
while the mediator decays into the corresponding SM fermion and the DM.

In general, the vector dark matter could either couple to the SM fermions democratically, or predominantly couple to leptons or quarks. 
As discussed in the literature~\cite{Agrawal:2011ze,Kile:2011mn,Batell:2011tc,Kamenik:2011nb}, the flavor structure of fermion portal dark matter could be very complicated. 
Similar to the sfermion as the fermion partner in the supersymmetric SM, every SM fermion could have one fermion partner in the fermion portal model.
As is known~\cite{Martin:1997ns}, the masses of the sfermions in the minimal supersymmetric SM follows the pattern that the first two generation partners are much heavier than the third generation partners. 
In our setup, to avoid the flavor constraints, we assume the similar mass spectrum for the mediators:
the fermion partners of the first two generation fermions are very heavy, while the partner masses of the third generation fermions could be sub-TeV. 
Thus, the partners of the first two generations are decoupled in the low energy simplified model. 
%
The mediator of the tau neutrino is absent if Majorana type neutrino is assumed.
Therefore, in the simplified model, we consider the DM exclusively interacts with three mediators: the $t'$, the $b'$ and the $\tau^\prime$, which are the fermion partners of the bottom quark, the tau lepton and the top quark respectively.
The relevant Lagrangian is written as 
\bea
	\mathcal{L}_{FfX} &=&  -  g_{Xb} \overline{{b'}} \gamma_\mu   P_R {b} X^\mu -  g_{X \tau} \overline{\tau^\prime} \gamma_\mu   P_R {\tau} X^\mu\nn\\
	&& -  g_{Xt} \overline{{t'}} \gamma_\mu   P_R {t} X^\mu + h.c..
\eea 
The above Lagrangian describes the fermion portal model. 
If the vector DM only couples to the third generation quarks, it is the quark portal model, while if the DM only couples to the tau lepton, it is the lepton portal model.

To explain the observed galactic center gamma-ray excess, we focus on the light vector dark matter with the mass range of $5-50$ GeV. In this case, the mass and coupling of the top quark partner become irrelevant in the relic density calculation and the indirect detection signature.
Therefore, we assume the same masses and couplings for the third generation quark partners $m_{t'} = m_{b'}$ and  $g_{Xt} = g_{Xb}$ without losing generality. 
Given the tight constraints on the masses of the third generation quark partners, we take $m_{t'} = m_{b'} = 900$ GeV as the benchmark point throughout the paper.
In the fermion portal sector, the new parameters are the tau partner mass $m_{\tau^\prime}$ and coupling $g_{X\tau}$, and the third generation quark partner mass $m_{b'}$ and coupling $g_{Xb}$. 
In the scalar sector, the independent parameters could be chosen as the scalar coupling $g_\phi$, the mixing angle $\varphi$ and the new scalar mass $m_S$. 
To avoid possible exclusion limits from the Higgs searches, we assume the scalar $S$ is heavier than the Higgs boson. 
If the new scalar $S$ is quite heavy, one could integrate out the heavy scalar, and obtain the effective theory at the low energy, which is the Higgs portal model.
In the following sections, we find that, due to very tight constraints from the direct detection experiments and the Higgs invisible decay, the parameter space in the scalar sector is very limited. 
Therefore, we expect that the scalar sector does not contribute to the relic density and indirect detection significantly. 
The dominant contribution to the relic density and to the indirect detection comes from the fermion portal sector.


\section{Thermal Relic Abundance}
\label{sec:relic}

The thermal relic abundance of the vector dark matter $X$ is governed by the 
Boltzmann equation of the dark matter number density $n_X$:
\bea
	\frac{{\rm d} n_X}{{\rm d} t} + 3 H n_X = - \langle \sigma v \rangle \left[n^2_X - n^2_{\rm EQ}\right],
\eea
where $H$ is the Hubble expansion rate, and $n^2_{\rm EQ}$ the number density at thermal equilibrium.
Here  $\langle \sigma v \rangle$ is the thermal averaged annihilation cross section times relative velocity.
In this study, depending on the dark matter mass, the vector dark matter could annihilate into the following final states: $f\overline{f}$, vector boson pair $VV$, di-Higgs $hh$. 
We focus on the light vector DM with mass less than $100$ GeV. For such light DM, the $hh$ and $t\bar{t}$ channels have not opened yet. 
Below the $W$ threshold, the dominant channels are $XX \to \tau\overline{\tau}$ and $XX \to b\overline{b}$, which comes from the $s$-channel scalar exchanges, and $t$-channel fermion portal. 
Above the $V$ boson thresholds, $XX \to VV$ channel through the $s$-channel scalar exchanges start to contribute to the relic density. 
In Appendix A, we present the annihilation cross section times relative velocity in each annihilation channel.

In the non-relativistic approximation, 
the annihilation cross section times relative velocity $\left(\sigma v\right)$ can be decomposed as 
\bea
\label{eq:xsecvab}
 \frac{\left(\sigma v\right)}{1 {\rm pb} \times c} = a + b v^2 + {\cal O}(v^4),
\eea
where the traditional unit is defined as
$1 {\rm pb} \times c  = 3 \times 10^{-26} {\rm cm}^3/{\rm s}$.
In the standard freeze-out approximation, the freeze-out value $x_f$ is obtain from solving the following equation iteratively~\cite{Jungman:1995df}  
\bea
	x^{1/2} e^{x} = c(c+2)\sqrt{\frac{45}{8}} \frac{g_i M_{\rm pl}}{2\pi^3} \frac{ m  \langle \sigma v \rangle }{\sqrt{g_{\rm{eff}}}} ,
\eea 
where  $M_{\rm pl}$ is the Planck mass, and the constant $c$ is usually taken to be $0.5$ in the calculation.
The thermal relic density is written as
\bea
 \Omega_{\rm DM} h^2 = \frac{m_X n_X}{\rho_{\rm crit}/h^2}  = \frac{ s_0 h^2}{\sqrt{\frac{\pi }{45}}\rho_{\rm crit}} \frac{1} {M_{\rm{pl}} \sqrt{g_{\rm eff}} {\mathcal I}(x_f) },
\eea
where $s_0$ is the entropy density of the present universe, $\rho_{\rm crit}$ the critical density. 
Here the function is defined as
\bea
	{\mathcal I}(x_f) = \int^{\infty}_{x_f} \frac{\langle \sigma v \rangle }{x^2} {\rm d}x.
\eea 
If we further take the non-relativistic approximation,  ${\mathcal I}(x_f)$ reduces to 
\bea
{\mathcal I}(x_f) = \frac{a + 3 b/x_f}{x_f}.
\eea
The current dark matter relic abundance has been measured by WMAP~\cite{Hinshaw:2012aka} and recently by Planck~\cite{Ade:2013zuv} with the following combined value 
\bea
	\Omega_{\rm dm} h^2 = 0.1199 \pm 0.0027.
\eea

We are interested in the light dark matter with mass around $5 \sim 50$ GeV, aiming at explaining the galactic center GeV gamma-ray excess.
In the mass range, the dominant channels are  $XX \to \tau\overline{\tau}$ and 
$XX \to b\overline{b}$.
The $t$-channel process gives the leading contribution to the $\tau\overline{\tau}$ and $b\overline{b}$ final states.
The $s$-wave component of the $t$-channel process $XX \to f\overline{f}$ depends on the dark matter mass and is not chirality-suppressed:
\bea \label{ffbar}
(\sigma v)_{\textrm{t-channel}} =  \frac{2 N_c g_X^4}{9 \pi } \frac{m_X^2}{(m_{X}^2 + m_F^2)^2} ,
\label{eq:relict}
\eea
where $N_c$ is the color factor of the SM fermion $f$, and $m_F$ the mass of the SM fermion partner.
On the other hand, the $s$-channel process is highly suppressed compared to the $t$-channel process. 
The $s$-wave component of the $s$-channel process is
\bea
(\sigma v)_{\textrm{s-channel}} & =& 
\frac{N_c m_f^2}{12\pi v^2} \frac{g_\phi^2 c_\varphi^2 s_\varphi^2 m_X^2 (m_S^2 - m_h^2)^2 }{(4m_X^2-m_h^2)^2 (4m_X^2-m_S^2)}. 
\label{eq:relics}
\eea
The above expression has the $m_f$ dependence, and is thus chirality-suppressed.
The $s$-channel thermal cross section is also proportional to $g_\phi  c_\varphi  s_\varphi$, which is highly constrained by the spin-independent direct detection experiments.
Moreover, there is no interference term between $t$-channel process and $s$-channel process due to the pure chiral coupling  of the vector DM to the fermion partner. 
Therefore, we expect that, for a light dark matter the $t$-channel contribution to the relic density is dominant over the $s$-channel one.

\begin{figure}[htp]
\begin{center}
\includegraphics[width=0.2\textwidth]{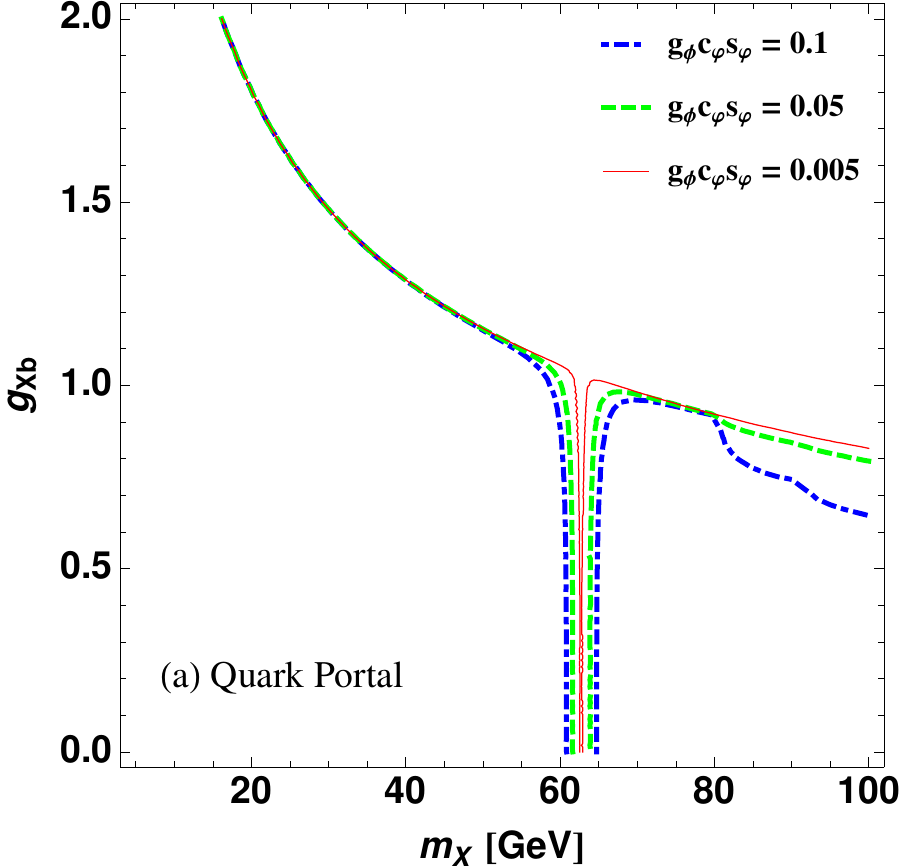} 
\includegraphics[width=0.2\textwidth]{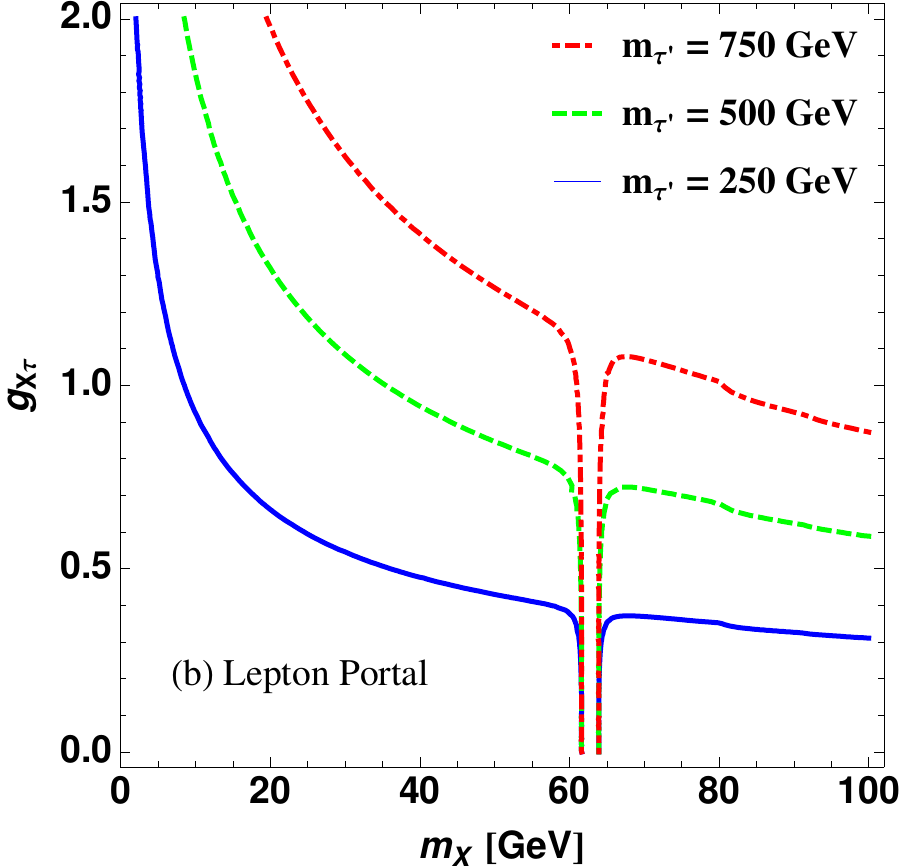} 
\caption{\small (a) The relic density contours for several scalar couplings $g_\phi c_\varphi s_\varphi$ in the dark matter mass $m_X$ verse  $g_{Xb}$ plane in the quark portal model. At the same time, we fix $m_{b'} = 900 $ GeV and $m_S = 500 $ GeV. 
(b) The relic density contours for several mediator masses $m_{\tau^\prime}$ in the dark matter mass $m_X$ verse   $g_{X\tau}$ plane in the lepton portal model. Here we take $g_\phi c_\varphi s_\varphi = 0.05$ and $m_S = 500 $ GeV.  }
\label{fig:relic1d}
\end{center}
\end{figure}

\begin{figure}[htp]
\begin{center}
\includegraphics[width=0.2\textwidth]{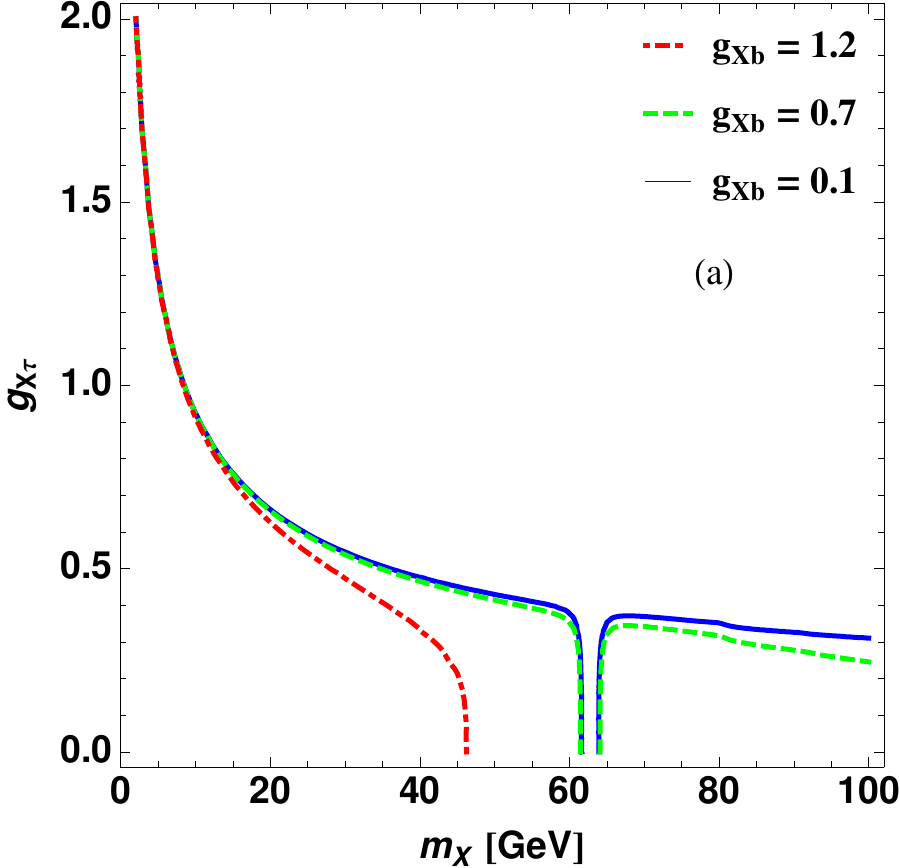} 
\includegraphics[width=0.2\textwidth]{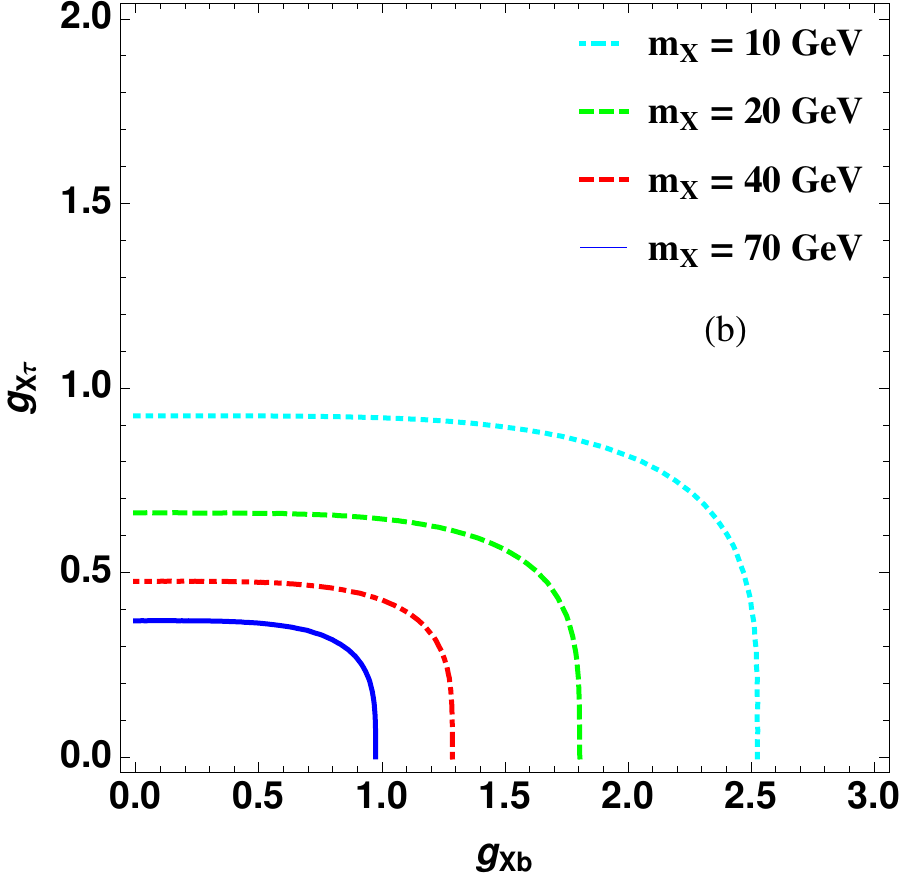} 
\caption{\small The relic density contours in the fermion portal model. (a) The contours of the $(m_X, g_{X\tau})$ plane for several $g_{Xb}$ couplings, (b) the contours of the $(g_{Xb}, g_{X\tau})$ plane for several dark matter masses $m_X$. Here the  mediator masses are fixed to be $m_{b'} = 900 $ GeV,$m_{\tau^\prime} = 250 $ GeV, and the scalar parameters are chosen to be $g_\phi c_\varphi s_\varphi = 0.05$ and $m_S = 500 $ GeV. }
\label{fig:relictb}
\end{center}
\end{figure}

From the thermally averaged cross sections in Eq.~\ref{eq:relict} and Eq.~\ref{eq:relics},
we note that the $t$-channel cross section is dependent on $g_X$ and $m_F$, while the $s$-channel cross 
section depends on the coupling $g_\phi c_\varphi s_\varphi$ and $m_S$, which are tightly constrained by the spin-independent direct detection and the Higgs invisible decay.
Fig.~\ref{fig:relic1d} (a) shows the $(m_X, g_X)$ contours which give the correct relic density  in the quark portal dark matter model.  
We fix the mediator mass $m_{b'} = 900 $ GeV, and illustrate how the $s$-channel contribution changes as we vary $g_\phi c_\varphi s_\varphi$. 
Our results show that the $s$-channel cross section contributes to less than 1\% of the relic density for the light dark matter.
In the Higgs resonance region, if the coupling $g_\phi c_\varphi s_\varphi$ is not so small, the $g_X$ coupling could be zero due to the enhancement of the thermal cross section near the Higgs resonance, as shown in Fig.~\ref{fig:relic1d}.
We also vary the scalar mass $m_S$ in the range of $250 \sim 1000$ GeV, and find that the relic density is insensitive to the scalar mass.
After knowing that the $s$-channel contribution is negligible,  we start to investigate the $m_F$-dependence in the dominant $t$-channel contribution. 
In Fig.~\ref{fig:relic1d} (b), we show the $(m_X, g_X)$ contours by fixing the scalar parameters and varying $m_F$. 
As expected, the relic density is quite sensitive to the fermion-portal parameters, due to the dominant contribution from the $t$-channel mediator.
In the fermion portal model, both the $b'$ and the $\tau^\prime$ mediator contribute to the relic density. 
We know that the $b'$ is tightly constrained by the LHC searches, while the constraint on the tau mediator is quite loose.
Therefore, we fix the $b'$ mass to the benchmark point $m_{b'} = 900 $ GeV, and vary the $\tau^\prime$ mass. 
In Fig.~\ref{fig:relic1d} (b), we study the dependence of the relic density to the $\tau^\prime$ mass.
Fig.~\ref{fig:relictb} (a) exhibits the $(m_X, g_{X\tau})$ contours for different $g_{Xb}$ with the mediator masses fixed.
It shows as the coupling strength $g_{Xb}$ increases, the $t$-channel $XX \to b \bar{b}$ dominates over the $t$-channel $XX \to \tau \bar{\tau}$. 
Fig.~\ref{fig:relictb} (b) shows the $(g_{Xb}, g_{X\tau})$ contours for several dark matter masses with mediator masses fixed. 
When the dark matter mass increases, the fermion-portal coupling strengths become smaller to obtain the correct relic density. 
%


\section{Direct Detection}

In the direct detection experiments, the WIMP usually scatters off target nucleus $T$
at low momentum transfer. 
At the zero momentum limit, the elastic scattering
cross section of the vector dark matter has the following spin-independent (SI) and the spin-dependent (SD) contributions:
\bea
\sigma_T =
\frac{\mu_T^2}{\pi} \bigg( \left| {\mathcal C}_p Z + {\mathcal C}_n (A-Z) \right|^2
+{\mathcal J}
|{\mathcal A}_p\langle S_p\rangle +{\mathcal A}_n\langle S_n\rangle|^2 \bigg),\nn\\	
\label{eq:ddeq}
\eea
where $\mu_T$ is the reduced mass of the nucleus-WIMP system, and the spin-induced factor ${\mathcal J}$ is
\bea
	{\mathcal J } = \frac{4j(j+1)}{3} \frac{4(J+1)}{J},
\eea
with $j$ the spin of the  dark matter, and $J$ the total spin of the target $T$, respectively.
Here $Z$ and $A$ are the atomic number and the weight of the target nucleus,
and $\langle S_{p,n}\rangle$ are the spin matrix element of the nucleus, and are taken from the Table I of the Ref.~\cite{Cannoni:2012jq}.  
${\mathcal C}_{p,n}$ and ${\mathcal A}_{p,n}$ are the effective WIMP-nucleon coupling strengths, which are model-dependent.  

\begin{figure}[htp]
\begin{center}
\includegraphics[width=0.18\textwidth]{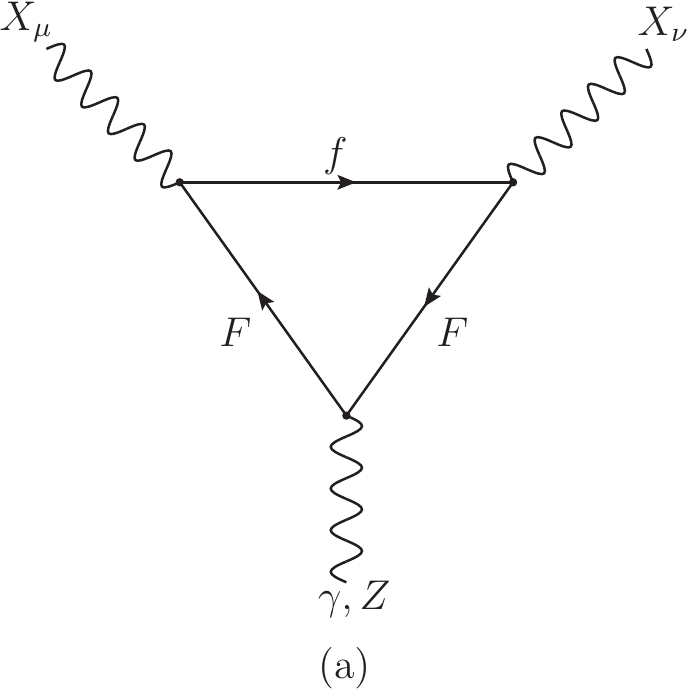} 
\includegraphics[width=0.18\textwidth]{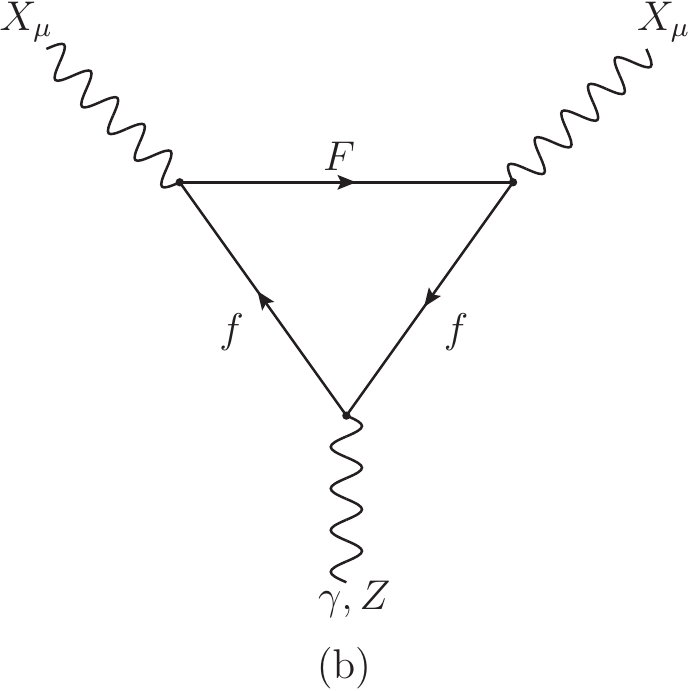} 
\caption{\small The one-loop effective trilinear vertex $\Gamma^{\mu\nu\rho}$ between the vector dark matter and the photon or the Z boson. The fermions appeared in the loop are the SM fermions $f$, and corresponding  mediators $F$. Here the mediators are  the $\tau^\prime$, the $b'$, and the $t'$. In Feynman diagram (a), the $F$ attaches to the photon and the Z boson, while $f$ attaches to the photon and the Z boson in (b). }
\label{fig:vvvtri}
\end{center}
\end{figure}

\begin{figure}[htp]
\begin{center}
\includegraphics[width=0.14\textwidth]{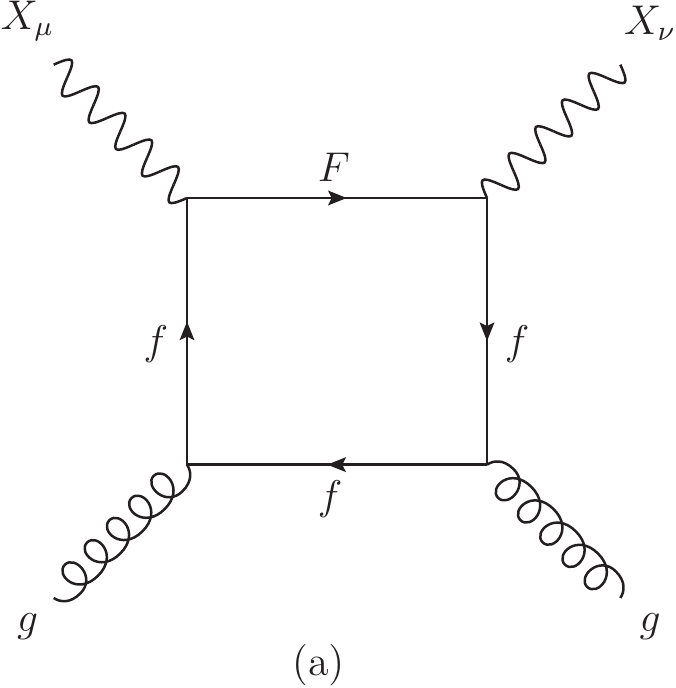} 
\includegraphics[width=0.14\textwidth]{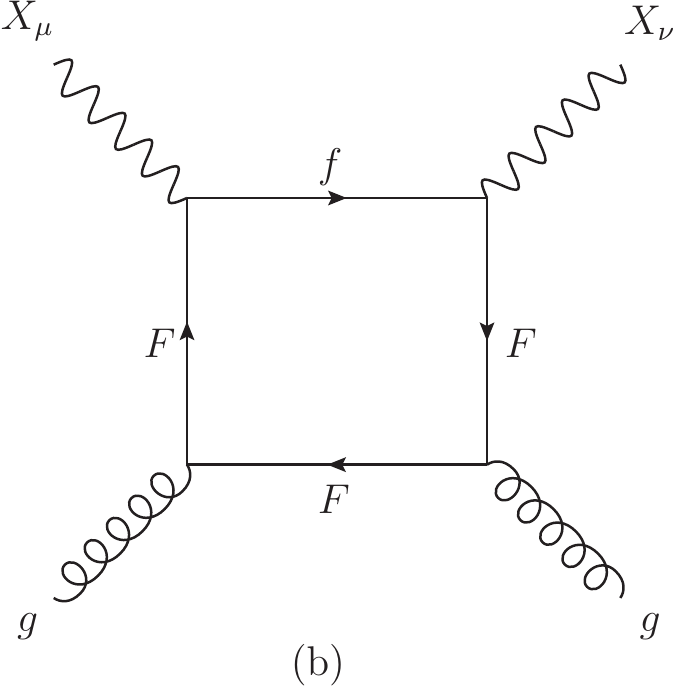} 
\includegraphics[width=0.14\textwidth]{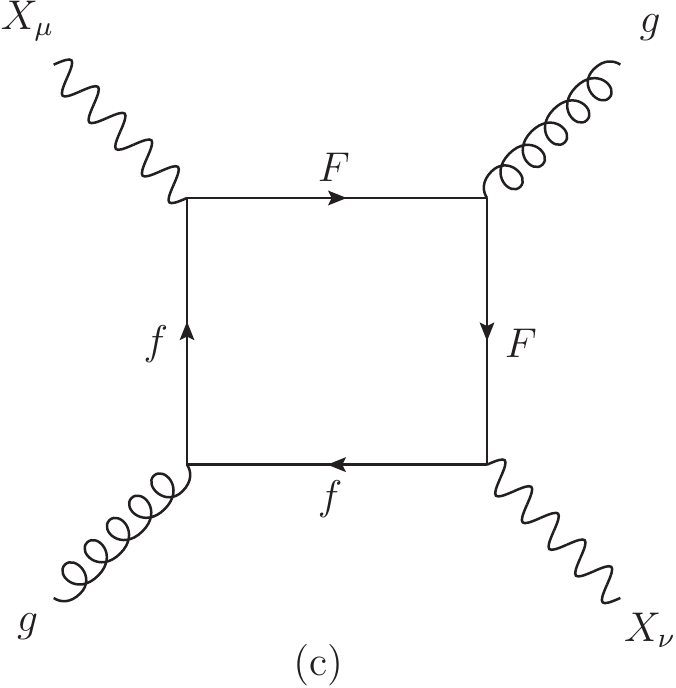} 
\caption{\small The one-loop Feynman diagrams (a-c) to show the vector dark matter scattering off the gluon in the direct detection.
The fermions appeared in the loop are the SM fermions $f$, and corresponding  mediators $F$.  Here the mediator $F$ can only be the $b'$, and the $t'$. }
\label{fig:vvvvbox}
\end{center}
\end{figure}

In our study, the DM particle not only interacts with the Higgs and the scalar, and also with the third generation fermions. 
The DM interactions with the Higgs and the scalar induce couplings of the DM to the nucleon. 
So the tree-level coupling of the WIMP to the nucleon is through scalar exchanges.  
The corresponding effective Lagrangian is 
\bea
	{\mathcal L} 
	&=& g_\phi c_\varphi s_\varphi m_X \frac{m_q}{v} \left( \frac{1}{m_h^2} - \frac{1}{m_S^2} \right) X_\mu X^\mu \overline{q} q\,.
	\label{eq:higgsLag}
\eea 
There are also non-negligible contributions at the one-loop level, from the DM interactions with the fermion mediators. 
The virtual photon triangle and box diagrams, as shown in Fig.~\ref{fig:vvvtri} and Fig.~\ref{fig:vvvvbox}, provide non-negligible contributions to the WIMP-nucleon scattering through effective couplings at one-loop order.
The contribution from the virtual $Z$ triangle diagrams is momentum suppressed due to the $\frac{q^2}{m_Z^2}$-dependence, and is thus sub-dominant. 
From the diagrams shown in Fig.~\ref{fig:vvvtri}, the resulting effective Lagrangian from the photon exchange is
\bea
	{\mathcal L} =  d_X \epsilon^{\mu\nu\rho\sigma} \left(X_\mu\partial_\nu X_\rho\right) \overline{q}\gamma_\sigma \gamma_5 q\,,
	\label{eq:sdtriloop}
\eea
where the coefficient $d_X$, in the zero external mass limit, is
\bea
	d_X  \simeq N_c \frac{eQ g_X^2 }{\pi^2}   \left( \frac{(m_F^6 - m_f^6)}{3(m_F^2 - m_f^2)^4} \log\frac{m_F}{m_f} - \frac{m_F^2 + m_f^2}{4(m_F^2 - m_f^2)^2} \right).\nn\\
\eea
The box diagrams shown in Fig.~\ref{fig:vvvvbox}, result in the following effective Lagrangian
\bea
	{\mathcal L} = b_g   X^{\rho} X_{\rho}G^{a\mu\nu}G^a_{\mu\nu},
	\label{eq:boxLag}
\eea
where the coefficient $b_g$, in the zero external mass limit, is~\cite{Hisano:2010yh}
\bea
	b_g =  \frac{\alpha_s g_X^2}{48\pi} \frac{3 m_F^2- 2m_X^2}{(m_F^2-m_X^2)^2}.
\eea
From the Appendix C, we find that the triangle loop diagrams contribute to the SD cross section, while the box loop diagrams
contribute to the SI cross section. 
The full results of the triangle and box diagrams are presented in Appendix D.

\begin{figure}[htp]
\begin{center}
\includegraphics[width=0.2\textwidth]{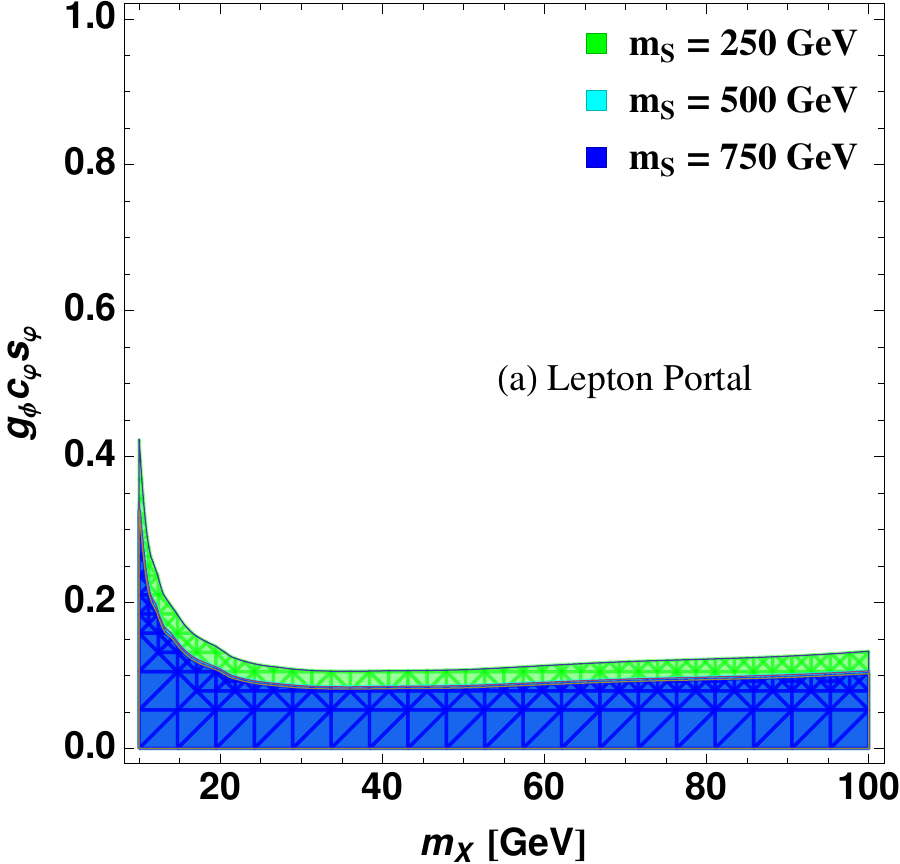} 
\includegraphics[width=0.2\textwidth]{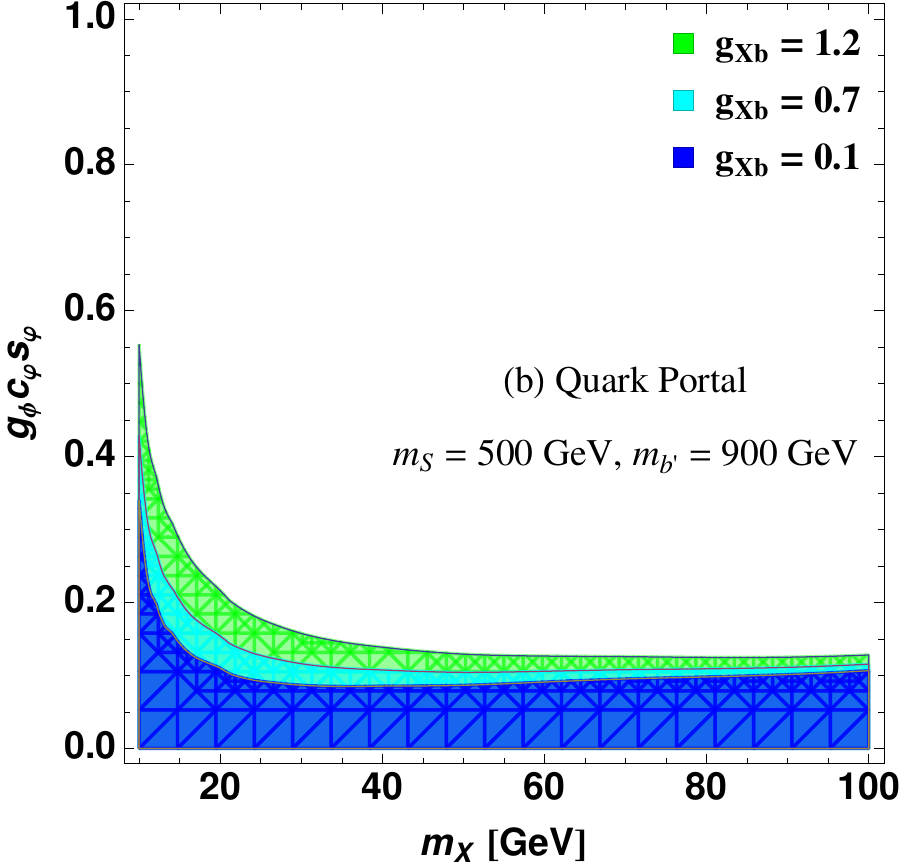} 
\caption{\small The allowed parameter spaces from the exclusion limits of the spin-independent cross section. (a) The allowed regions of the coupling $g_\phi c_\varphi s_\varphi$ verse the dark matter mass $m_X$ plane, for three different values of the scalar masses, in the lepton portal model. (b) The allowed regions of the coupling $g_\phi c_\varphi s_\varphi$ verse the dark matter mass $m_X$ plane, for three different values of the couplings $g_{Xb}$, in the quark portal model. We choose $m_S = 500 $ GeV and $m_{b'} = 900 $ GeV to fix the box-diagram contributions in the quark portal model. The allowed region in the fermion portal model is the same as the one in (b).
}
\label{fig:ddsi}
\end{center}
\end{figure}

\begin{figure}[htp]
\begin{center}
\includegraphics[width=0.2\textwidth]{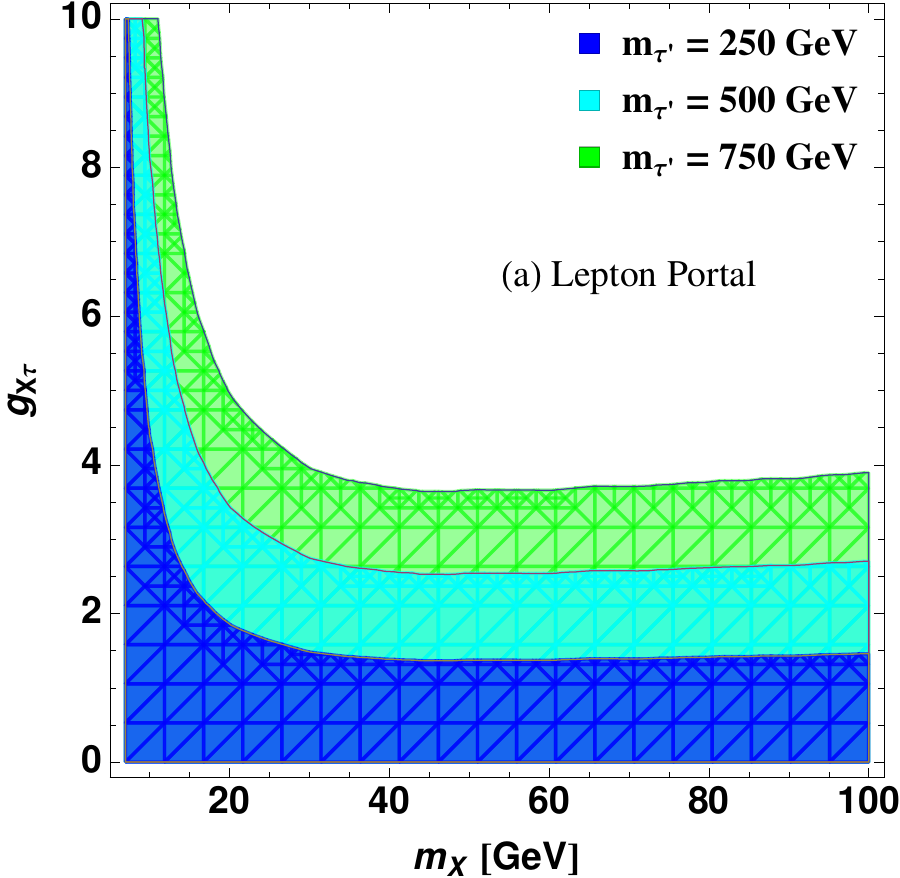} 
\includegraphics[width=0.2\textwidth]{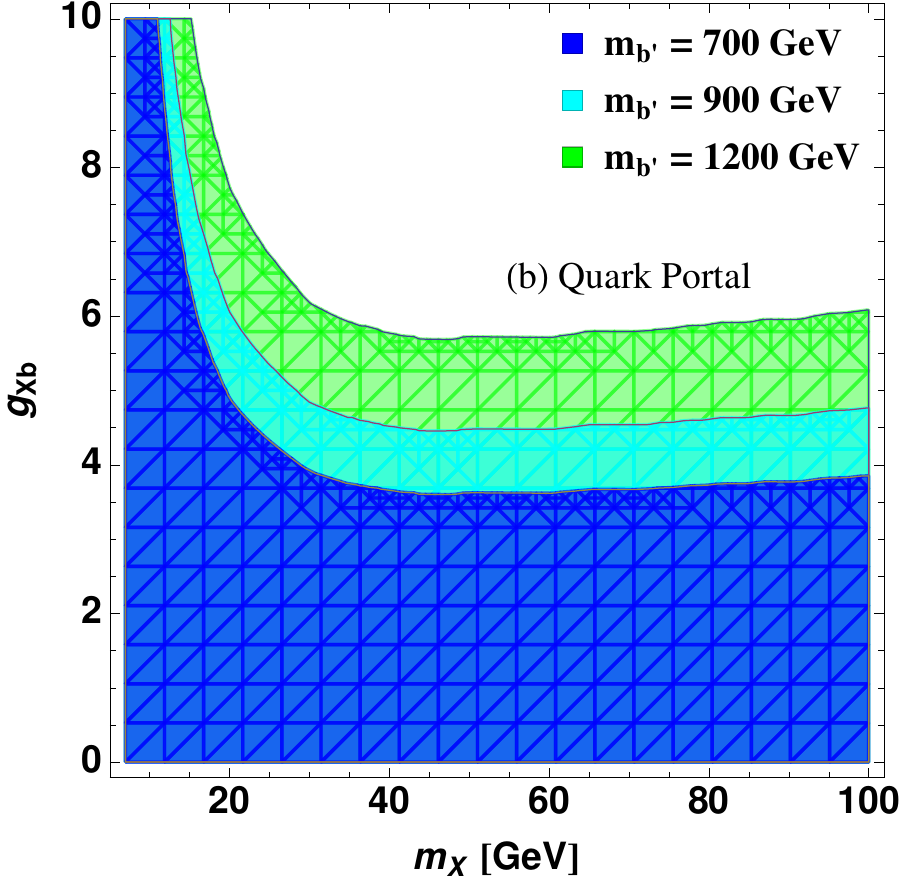} 
\caption{\small  The allowed parameter spaces from the exclusion limits of the spin-dependent cross section. (a) The allowed regions of the $(m_X, g_{X\tau})$ plane for several mediator masses in the lepton portal model,  and (b) the $(m_X, g_{Xb})$ plane  for several mediator masses in the quark portal model. }
\label{fig:ddsd}
\end{center}
\end{figure}

\begin{figure}[htp]
\begin{center}
\includegraphics[width=0.2\textwidth]{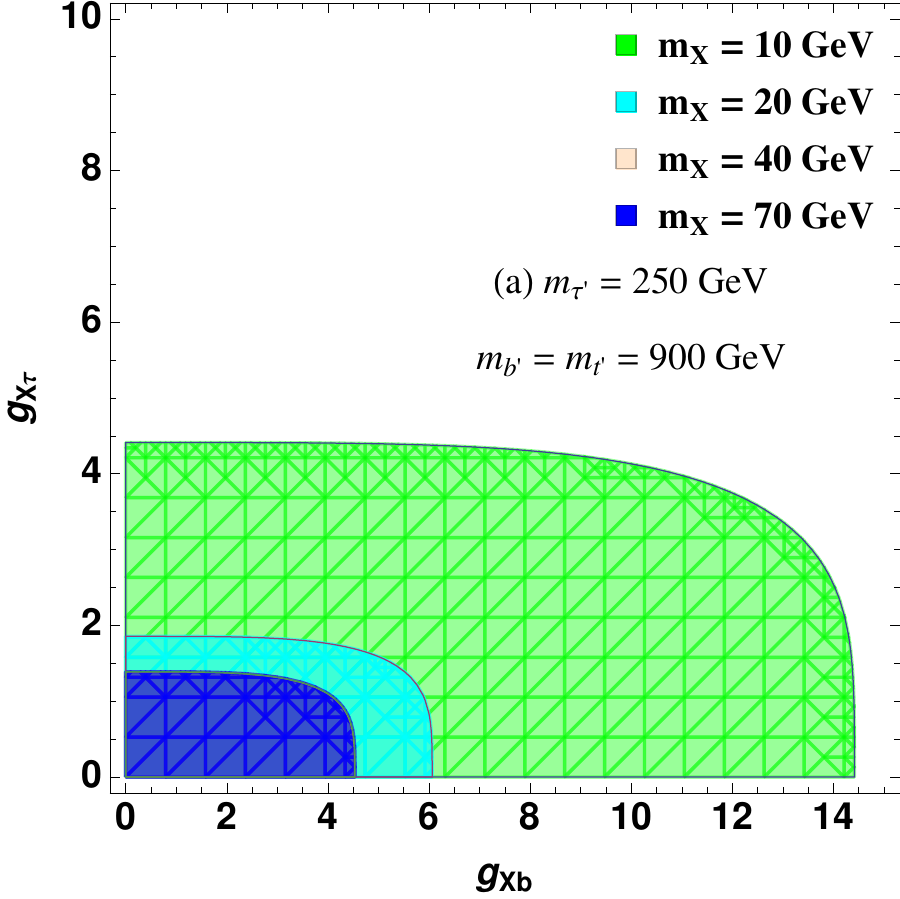} 
\includegraphics[width=0.2\textwidth]{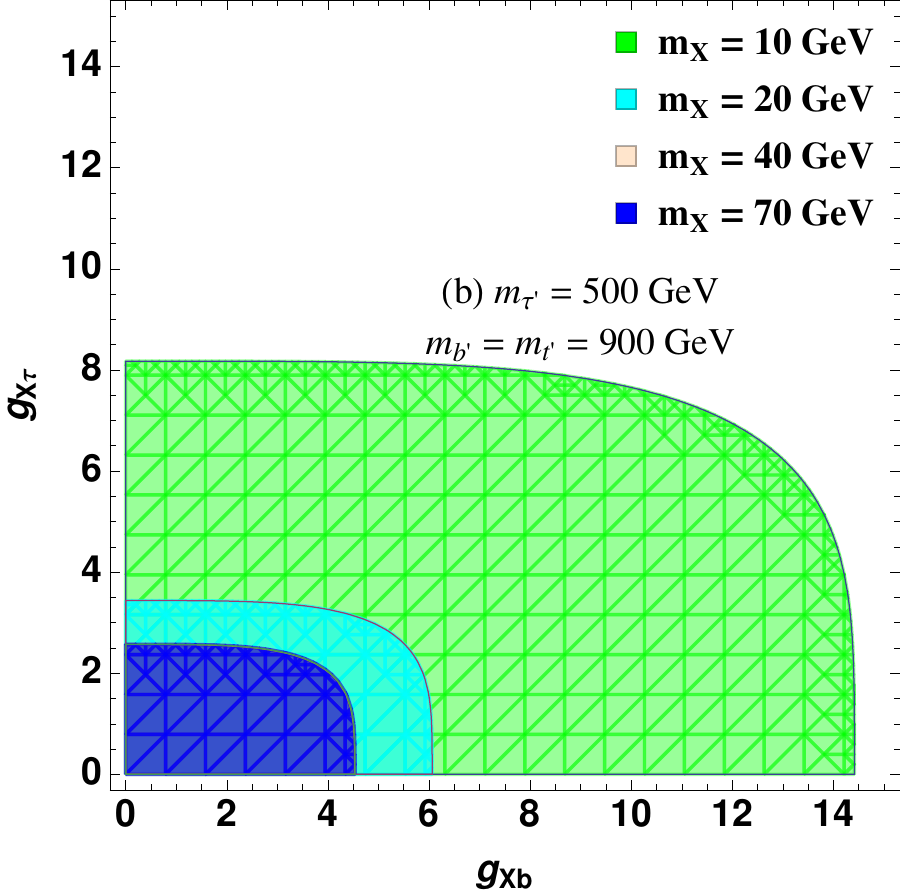} 
\caption{\small The allowed regions of the $(g_{Xb}, g_{X\tau})$ plane from the exclusion limits of the spin-dependent cross section in the fermion portal model. Here we take the benchmark points: $m_{\tilde{b}} = m_{\tilde{t}} = 900 $ GeV, $m_{\tilde{\tau}} = 250 $ GeV in (a), and $m_{\tilde{\tau}} = 500 $ GeV in (b). }
\label{fig:ddsdtb}
\end{center}
\end{figure}

To obtain the elastic cross section, we connect the effective Lagrangian at the quark and gluon level with the matrix element at nucleon level, discussed in Appendix C. 
Given the effective Lagrangian in Eqs.~\ref{eq:higgsLag} and~\ref{eq:boxLag}, we obtain the SI coefficient ${\mathcal C}_{p,n}$ defined in Eq.~\ref{eq:ddeq}:
\bea
	{\mathcal C}_{N} &=& \frac{g_\phi  c_\varphi s_\varphi}{2}  \frac{m_N}{v} \left(1- \frac79 f^{(N)}_{TG}\right) 
	\left( \frac{1}{m_h^2} -  \frac{1}{m_S^2} \right) \nn\\
	&&- \frac{g_X^2}{108}f^{(N)}_{TG} \frac{m_N}{m_X}\frac{3 m_F^2- 2m_X^2}{(m_F^2-m_X^2)^2}  ,
	\label{eq:si_coup}
\eea
where $f^{(N)}_{TG}$ is defined in Appendix C.
%
%
The effective Lagrangian in Eq.~\ref{eq:sdtriloop} contributes to the SD cross section.
The $d_X$ term contributes to 
\bea
	{\mathcal A}_{N} = e d_X \sum_{q} \Delta^N_q,
\eea
where $e \Delta^N_q$ comes from the axial vector coupling of the quarks.
The final results of 
the elastic cross sections are expressed as
\bea
	\sigma^{\rm SI}_N &=&  \frac{\mu_N^2}{\pi} {\mathcal C}_{N}^2, \nn\\
	\sigma^{\rm SD}_N &=&  \frac{16\mu_N^2}{\pi} {\mathcal A}_{N}^2.
\eea

The most stringent limits on the DM coupling to the nucleon are set by LUX~\cite{Akerib:2013tjd}, XENON100~\cite{Aprile:2012nq}, SuperCDMS~\cite{Agnese:2014aze} and CRESST~\cite{Angloher:2014myn} for the SI interactions.  
In the lepton portal model, only the scalar exchanges contribute to the SI cross section. 
Given the LUX exclusion limits on the nucleon cross section, we obtain very tight constraint on the scalar coupling:  $g_\phi c_\varphi s_\varphi < 0.1$ for the dark matter with mass $20 - 100 $ GeV, as shown in Fig.~\ref{fig:ddsi} (a). 
%
Fig.~\ref{fig:ddsi} (a) also implies that the scalar mass is insensitive to the direct detection bounds.
In the quark portal model, both the scalar exchanges and the box diagrams contribute to the SI cross section.
In Fig.~\ref{fig:ddsi} (b), we show the $(m_X, g_\phi c_\varphi s_\varphi)$ contours for several values of the $g_{Xb}$ at $m_S = 500 $ GeV.
The box diagrams are sub-dominant compared to the scalar exchanges.
Thus the allowed contour does not change much when varying the coupling $g_{Xb}$.
In the fermion portal model, the SI cross section is the same as the one in the quark portal model. 
The allowed parameter space is the same as the one shown in  Fig.~\ref{fig:ddsi} (b).  
Therefore, the exclusion limits on the SI cross section puts very strong constraint on the parameters in the scalar sector, but not the fermion portal sector.

For the SD interactions, PICASSO~\cite{Archambault:2012pm}, SIMPLE~\cite{Felizardo:2011uw} and COUPP~\cite{Behnke:2012ys} set bounds on DM-proton couplings, while XENON100~\cite{Aprile:2013doa} places constraints on the DM-neutron couplings.
The current exclusion limits on the SD cross section are much weaker than the tightest limits on the SI cross section. 
From Eq.~\ref{eq:sdtriloop}, only the triangle loop diagrams contribute to the SD interactions. 
We apply the COUPP and XENON100 exclusion limits and put constraints on the fermion portal parameters.  
Fig.~\ref{fig:ddsd} (a) and (b) show the allowed contours in the $(m_X, g_{X\tau})$ plane in the lepton portal model, and the $(m_X, g_{Xb})$ plane in the quark portal model, for several values of the  mediator masses.  
The results show that the coupling strength of the $g_{X\tau}$ and $g_{Xb}$ greater than 2 are allowed, due to the loop suppression of the triangle diagrams. 
In Fig.~\ref{fig:ddsdtb} (a) and (b), we present the $(g_{Xb}, g_{X\tau})$ contours for several dark matter masses in the fermion portal model.
It is interesting to see that the heavier of the dark matter, the tighter constraints on the couplings. 
The results show that still a large region of the fermion portal parameters is allowed by the current direct detection experiments.


\section{Electroweak Measurements and Higgs Invisible Decay}

\begin{figure}[htp]
\begin{center}
\includegraphics[width=0.22\textwidth]{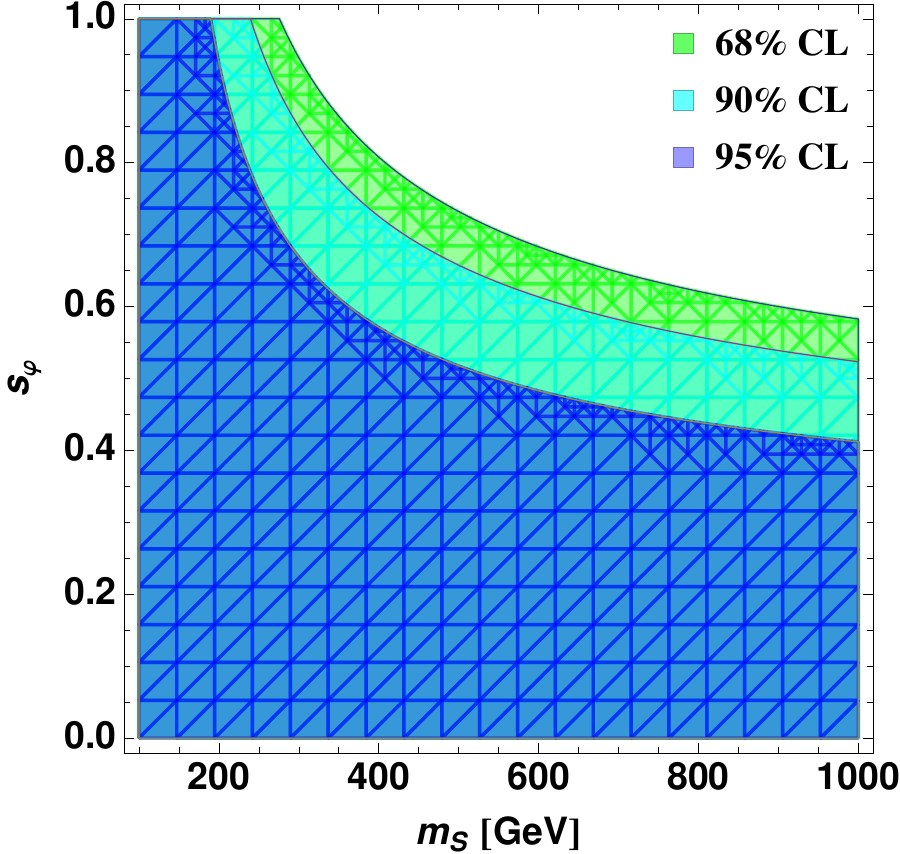} 
\includegraphics[width=0.22\textwidth]{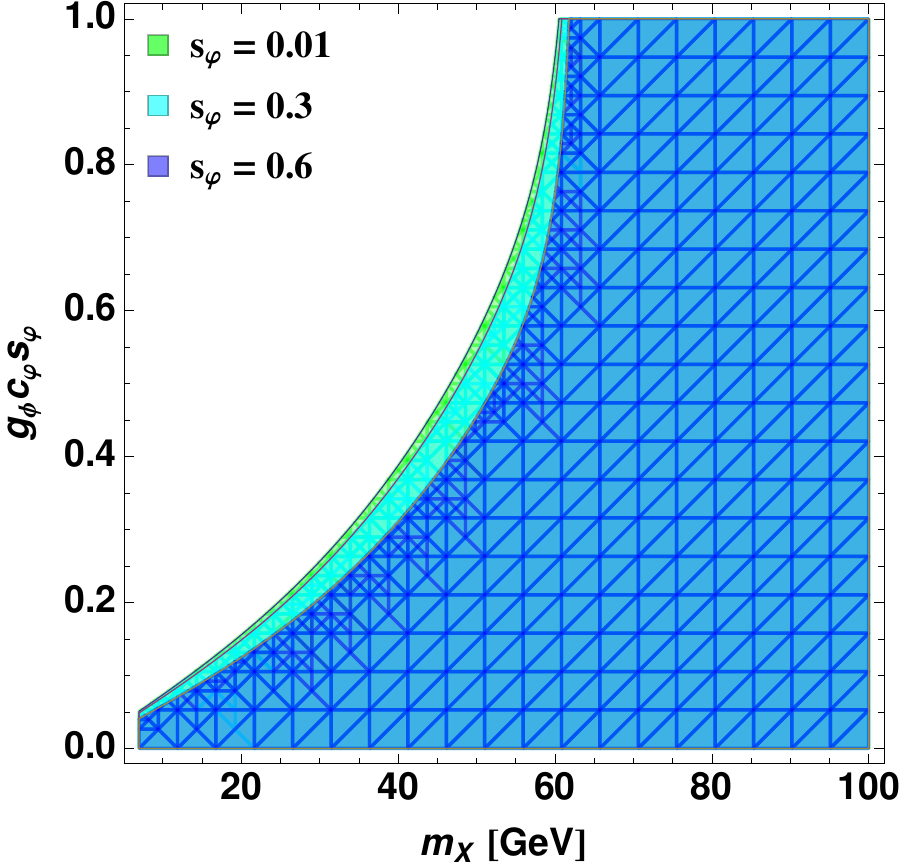} 
\caption{\small Left: the allowed regions of the $(m_S, s_\varphi)$ plane at the 68\%, 90\%, and 95\% confidence levels. Right: the allowed parameter space as a function of  $m_X$ and $g_\phi c_\varphi s_\varphi$ for three values of the mixing angle $s_\varphi$.  }
\label{fig:higgs}
\end{center}
\end{figure}

The electroweak observables, precisely measured at the LEP and SLC, put constraints on the model parameters. 
The dominant NP effects on the electroweak observables are the oblique corrections $S, T$~\cite{Peskin:1991sw}  in the gauge boson vacuum polarization correlation.
In the fermion sector, due to absence of the mixing between the SM fermions and their partners, 
the contribution from the SM fermion partners is zero. 
On the other hand, the mixing between the Higgs boson and new scalar exists in the scalar sector.
Hence, the electroweak precision data only constrain the parameters in the scalar sector.
It is straightforward~\cite{Xiao:2014kba} to calculate the oblique corrections due to the Higgs boson and the new scalar:
\bea
\Delta T  &=&  s_{\varphi}^2 \Big[ T_s(m_S^2) - T_s(m_h^2)  \Big], \\
\Delta S  &=&  s_{\varphi}^2 \Big[ S_s(m_S^2) - S_s(m_h^2)  \Big],
\eea
where the functions are defined as
\bea
\label{eq:oblqf}
T_s(m)&=&-\frac{3}{16\pi c_W^2}\bigg[\frac{1}{(m^2-m_Z^2)(m^2-m_W^2)}\notag\\
&&\times\Big(m^4\ln m^2-s_W^{-2}(m^2-m_W^2)m_Z^2\ln m_Z^2\notag\\
&+&s_W^{-2}c_W^2(m^2-m_Z^2)m_W^2\ln m_W^2\Big)-\frac{5}{6}\bigg], \\
S_s(m)&=&\frac{1}{12\pi}\bigg[\ln m^2-\frac{(4m^2+6m_Z^2)m_Z^2}{(m^2-m_Z^2)^2}\notag\\
&+&\frac{(9m^2+m_Z^2)m_Z^4}{(m^2-m_Z^2)^3}\ln\frac{m^2}{m_Z^2}-\frac{5}{6}\bigg].
\eea
Note that if the mixing angle $s_\varphi$ goes to zero, there is no oblique correction. 
In Fig.~\ref{fig:higgs} (left), we show the allowed parameter space as a function of the mixing angle and the new scalar mass.
The constraints on the mixing angle and the new scalar mass from the electroweak data are quite weak as expected.
The reason is that the dominant contribution from the new scalar has $\ln m_S^2$ dependence, which is  similar to the Higgs contributions in the SM.

After the Higgs is discovered, the Higgs couplings and its width are measured precisely at the LHC. 
If the dark matter mass is less than half of the Higgs boson mass, the Higgs will decay invisibly to $h \to XX$.
Both ATLAS and CMS~\cite{Aad:2014iia,Chatrchyan:2014tja} set upper limits on the Higgs invisible width.
In our model, the invisible decay width is
\bea
	\Gamma_h^{\rm inv} &=& \frac{g_{XXh}^2}{64 \pi} \frac{m_h^3}{m_X^4} \left( 1  - \frac{4 m_X^2}{m_h^2} + \frac{12 m_X^4}{m_h^4}  \right)  \left( 1 - \frac{4 m_X^2}{m_h^2} \right)^{1/2}
\eea
This puts very strong constraints on the scalar sector if the dark matter is very light.
For a dark matter with mass close to half of the Higgs boson mass, the constraints become weak due to the kinematic suppression.
In Fig.~\ref{fig:higgs}(right), we show the allowed parameter space as a function of the the mixing angle and the new scalar mass from the invisible Higgs decay.
When the dark matter is very light, the coupling $g_\phi c_\varphi s_\varphi$ should be very small. 
This is complementary to the SI direct detection constraints on the coupling $g_\phi c_\varphi s_\varphi$. 
Therefore, the constrains from the SI direct detection and Higgs invisible decay exclude the region where the   coupling $g_\phi c_\varphi s_\varphi$ is greater than 0.1 for the whole region of the dark matter mass.


\section{LHC Searches on Fermion Mediators}

The mediators have the same charge and color quantum numbers as the SM fermions. 
So if masses of some mediators  are sub-TeV, those can be produced on-shell in pair and then subsequently decay to the vector dark matter and 
corresponding SM fermions, resulting in the final states with the SM fermion pairs and the transverse missing energy (MET).
In our setup, only mediators of the third-generation fermions are not so heavy.  
So we only consider possible signatures of three mediators: the heavy bottom $b'$, and the heavy top $t'$, the heavy tau $\tau^\prime$.  
The signatures at the LHC are the bottom quark pair plus MET, the top pair plus MET, and the tau lepton pair plus MET, separately.

\begin{figure}[htp]
\begin{center}
\includegraphics[width=0.22\textwidth]{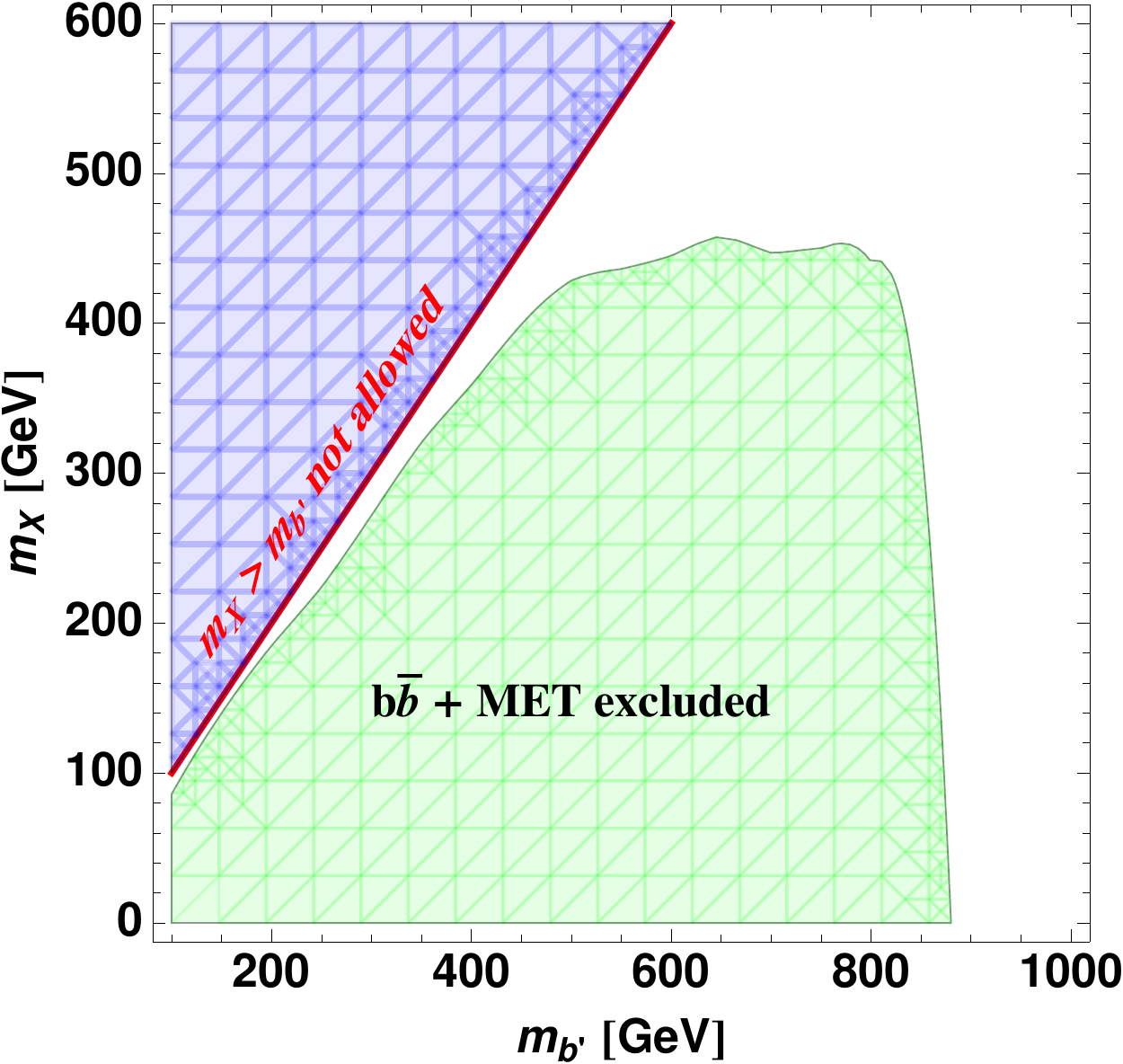} 
\includegraphics[width=0.22\textwidth]{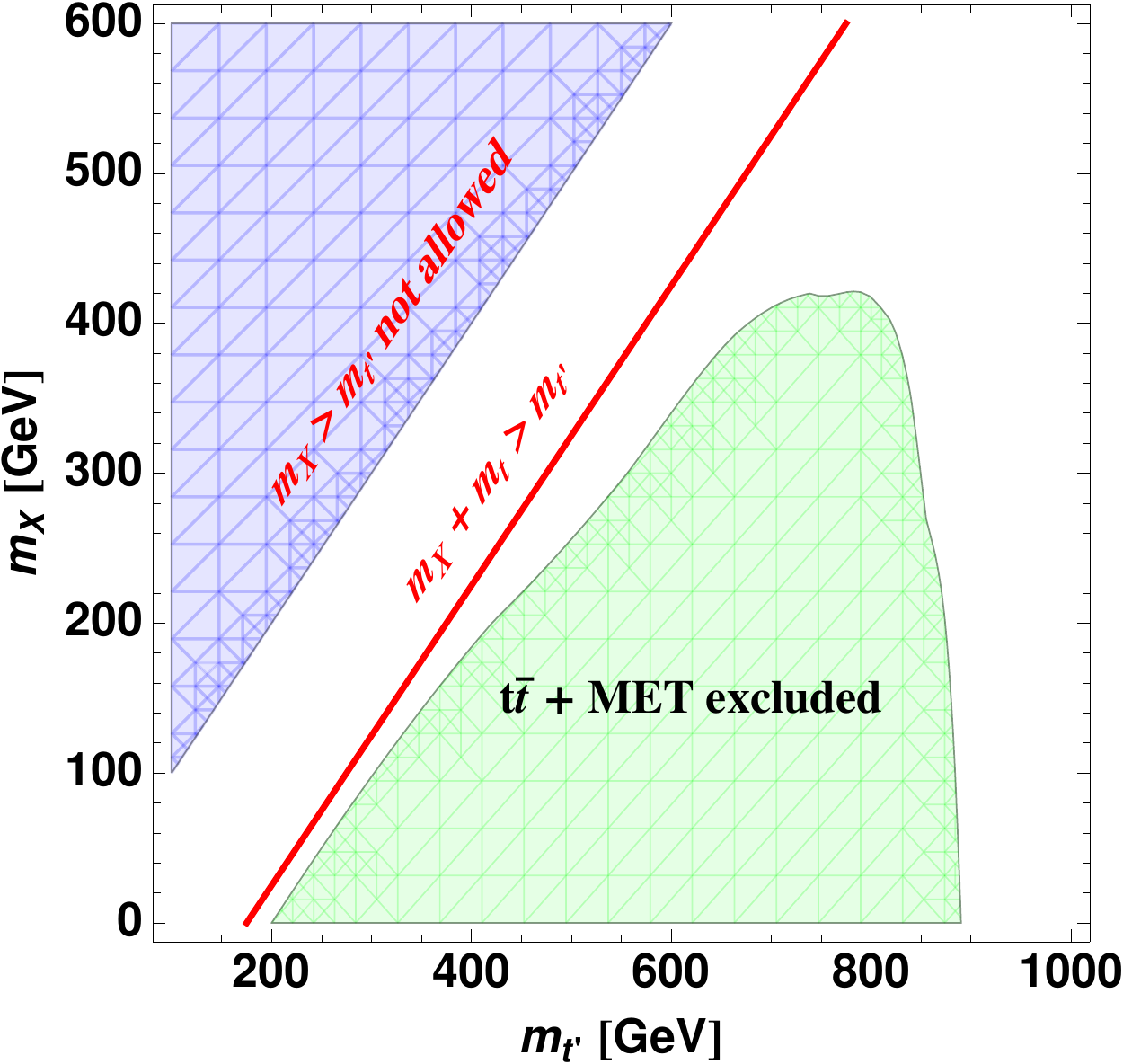} 
\caption{\small Left: The excluded regions on the parameter space $(m_{b'}, m_X)$ from LHC searches on the the bottom pair plus missing transverse energy final states with $20.1 $ fb$^{-1}$ luminosity at the 8 TeV. 
Right: The excluded regions on the parameter space $(m_{t'}, m_X)$ from LHC searches on the the hadronic top pair plus missing transverse energy final states with $20.3 $ fb$^{-1}$ luminosity at the 8 TeV.  }
\label{fig:lhc}
\end{center}
\end{figure}

These final states have been investigated by both the ATLAS and the CMS~\cite{Aad:2013ija,CMS:2014nia,Aad:2014bva,Aad:2014kra,Aad:2014qaa,CMS:2014wsa}. In these analyses, both the ATLAS and the CMS utilize the final states to set the exclusion limits on the sbottom quark, the stop quark and sleptons.
Since the searches on sfermions and the mediators share the same event topology, 
the existing search limits on the sfermions could be translated into the search limits on the fermion portal mediators.
However, the production cross sections of the sfermions are different from the ones of the mediators. 
Therefore, we calculated the next-to-next-to-leading-order (NNLO) cross section of the heavy top and bottom mediators using the Hathor package~\cite{Aliev:2010zk}. 
We use the most updated and stringent search limit: ATLAS analysis on sbottom quark searches~\cite{Aad:2013ija} with $20.1 $ fb$^{-1}$ luminosity at the 8 TeV, and ATLAS analysis on stop quark searches~\cite{Aad:2014bva} with $20.3 $ fb$^{-1}$ luminosity at the 8 TeV. 
There is no existing exclusion limit on the stau from the LHC, 
although ATLAS investigate the tau lepton pair plus MET final states~\cite{Aad:2014yka}. 
The current constraints on the stau still comes from the LEP data~\cite{PDG2014}.
Therefore, the converted constraints on the $\tau^\prime$ is $m_{\tau^\prime} > 81.9$ GeV.
After taking care of the difference in the production cross sections, we convert the exclusion limits on the squarks into the exclusion limits on the mediators.
In Fig.~\ref{fig:lhc}, we present our translated exclusion limits on the heavy top and bottom mediators. 
It shows that the limits on the quark mediators are $m_{t'}, m_{b'} > 890 $ GeV, irrelevant to the coupling $g_{Xb}$ in the quark portal sector. 
This exclusion limits put the tightest constraints on the quark portal sector.

The current tight limits on the  $t', b'$ masses surpass the flavor constraints. 
Because $t', b'$ are $Z_2$-odd particles, the most sensitive observables in flavor physics comes from the $B_d-\overline{B}_d$ and  $B_s-\overline{B}_s$ systems~\cite{PDG2014}. 
The box diagrams involving two $X_\mu$ and two $b'$ contribute to the B meson mixing. 
Ref.~\cite{Hubisz:2005bd} calculated the box contributions and found that the results of the box diagrams 
depend on the suppression factor  $\frac{m_X^4}{m_{b'}^4}$. 
Unlike the enhancement factor $\frac{m_t^2}{m_W^2}$ in the box diagrams with the top quark and charged W boson involved,
the box diagrams involving the $b'$ and neutral gauge boson $X_\mu$ are highly suppressed by the $b'$ mass. 
If we apply $m_{b'} > 890 $ GeV, the flavor constraints on the $b'$ couplings are quite weak. 
Therefore, we neglect the limits from the flavor physics.

One may expect the mono-jet searches at the LHC could also constrain on the quark portal sector.
This fermion portal model predicts the mono-bottom jet and mono-top jet final states at the LHC.
On the other hand, the current searches at the LHC~\cite{Khachatryan:2014rra,ATLAS:2012zim}
focus on the mono-light jet final states. 
Therefore, we expect that if we apply the exclusion limit for the mono-light jet searches, 
the converted exclusion limit will be quite weak, and thus  cannot compete the limits from the direct production searches.
%


\section{Indirect Detection: Gamma-Ray Excess at Galactic Center}

\begin{figure}[htp]
\begin{center}
\includegraphics[width=0.36\textwidth]{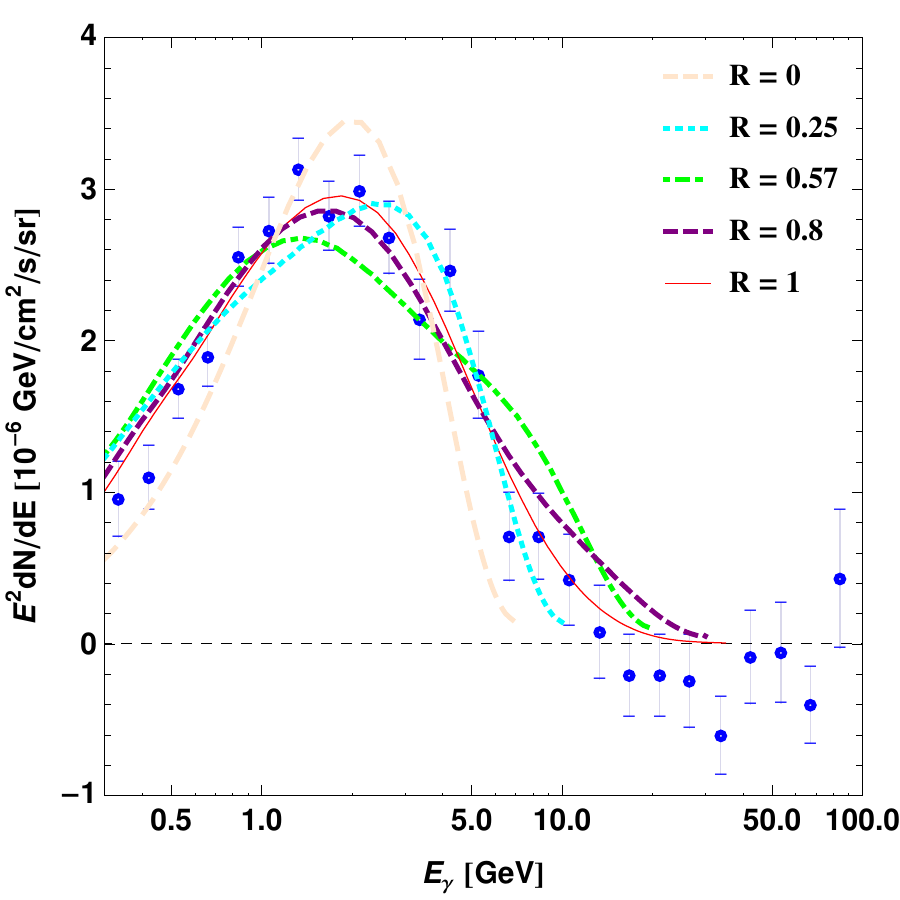} 
\caption{\small Data of the gamma-ray flux observed in the inner galaxy, taken from Ref.~\cite{Daylan:2014rsa}.  The curves are the best-fit spectra for several relative weight $R$, which parametrizes the combination of the $b\bar{b}$ and $\tau\bar{\tau}$ final states. For each curve, the flux is calculated with a angular direction $5^{\circ}$ from the galactic center, and a generalized NFW profile is used. }
\label{fig:flux}
\end{center}
\end{figure}

To explain the observed gamma-ray excess in the inner region of our galaxy in this fermion portal model,
we study the gamma-ray spectrum and flux from the vector dark matter annihilating into the $b\bar{b}$ and $\tau\bar{\tau}$ final states.
In general, the differential flux of gamma-rays from a given angular region $\Delta \Omega$ is given by
\bea
	\frac{{\rm d} \Phi^\gamma}{{\rm d} E_\gamma} = \frac{1}{8\pi m_X^2}  \langle \sigma v\rangle 
	\frac{{\rm d} N^\gamma}{{\rm d} E_\gamma} \times \int_{\Delta \Omega} J(\psi) {\rm d}\Omega,
\eea
where $\psi$ is the angle from the direction of the Galactic Center that is observed, $\langle \sigma v\rangle$ is the total thermally averaged cross section,
and $\frac{{\rm d} N^\gamma}{{\rm d} E_\gamma}$ is the gamma-ray spectrum produced per annihilation. 
The $J$ factor is obtained from the line of sight integration~\cite{Cirelli:2010xx}
\bea
	J(\psi) = \int_{\rm l.o.s.} [\rho(r(s,\psi))]^2 {\rm d} s,
\eea
where $\rho(r(s,\psi))$ is the dark matter halo profile, $r(s,\psi) = \sqrt{r^2_\odot+s^2 - 2 r_\odot s \cos\psi}$ with $r_\odot = 8.5 $ kpc. The dark matter halo profile is well-fitted by a generalised Navarro-Frenk-White (gNFW) distribution~\cite{Navarro:1995iw,Navarro:1996gj}
\bea
	\rho(r)=\rho_0\frac{(r/r_s)^{-\gamma}}{(1+r/r_s)^{3-\gamma}}~,
\eea
where $\rho_0$ is selected to have the local dark matter density $0.3~{\rm GeV}/{\rm cm}^3$ at the distance  $r_\odot$ , the scale radius of 
$r_s$ is adopted to be $20$ kpc,
and the best-fit value for the slope of the gNFW profile $\gamma$ is $1.26$.

\begin{figure}[htp]
\begin{center}
\includegraphics[width=0.22\textwidth]{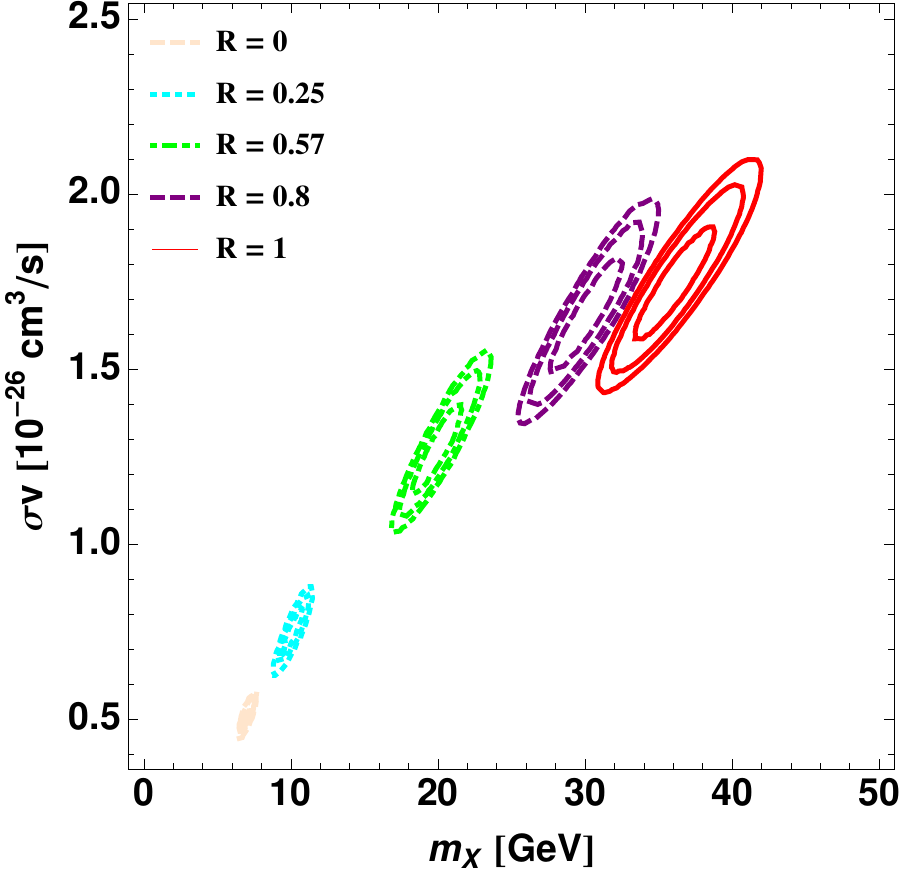} 
\includegraphics[width=0.22\textwidth]{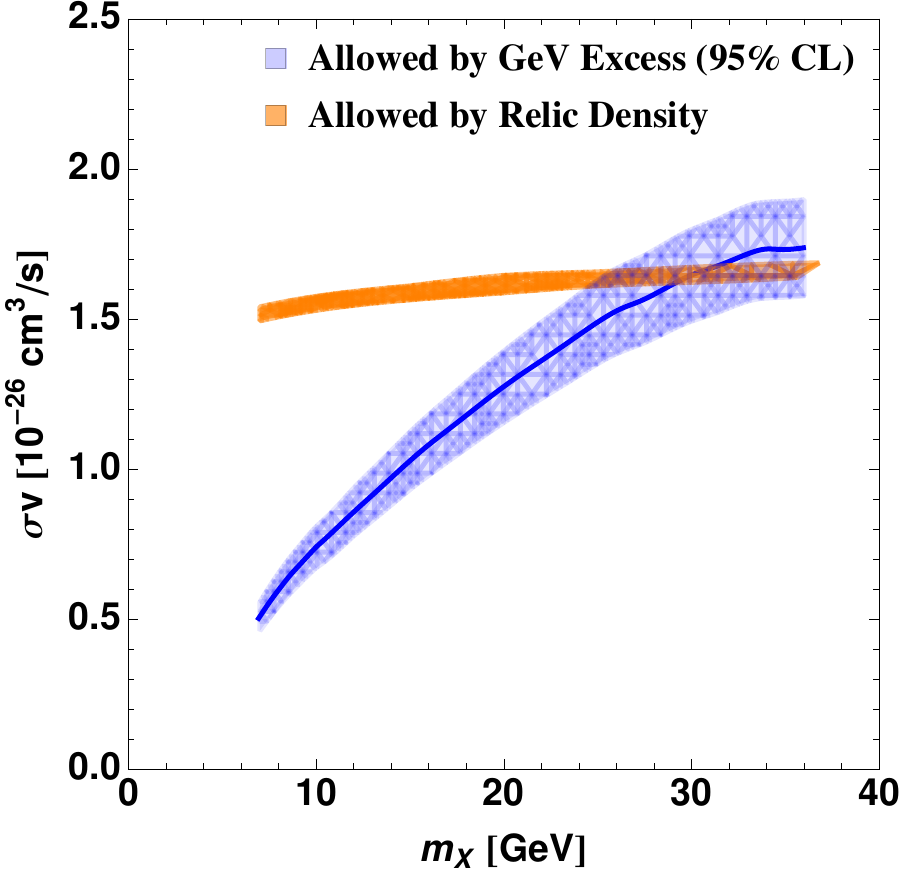} 
\caption{\small Left: The contours of the dark matter mass and annihilation cross section required to  fit of the gamma-ray spectrum at the 68\%, 90\%, and 95\% confidence levels, for a variety of the relative weight $R$. 
Right: The regions of the annihilation cross section at the 95\% confidence level verse the best fit of the dark matter mass, compared to the allowed region by the requirement of the correct relic density.   }
\label{fig:chi}
\end{center}
\end{figure}

Since the vector dark matter could annihilate into the $b\bar{b}$ and $\tau\bar{\tau}$ final states, 
we study which combinations of the dark matter mass and SM final states provide a good fit to the gamma-ray flux.
For this purpose, we introduce a ratio $R$ to parametrize the relative weight of the $b\bar{b}$ and $\tau\bar{\tau}$ final states. 
Using the ratio $R$, the gamma-ray spectra produced per annihilation can be written as 
\bea
	\frac{{\rm d} N^\gamma}{{\rm d} E_\gamma} = R\ \frac{{\rm d} N^\gamma_{b\bar{b}}}{{\rm d} E_\gamma}  + (1-R)\ \frac{{\rm d} N^\gamma_{\tau\bar{\tau}}}{{\rm d} E_\gamma}.
	\label{eq:flux}
\eea
In the limit $R \to 1$, we recover the gamma-ray spectrum from dark matter annihilating into $b\bar{b}$ final state in the quark portal model,   
while  $R \to 0$ in the lepton portal model.
For the two-body annihilation, the gamma-ray spectra $\frac{{\rm d} N^\gamma_{b\bar{b}}}{{\rm d} E_\gamma}$ and 
$\frac{{\rm d} N^\gamma_{\tau\bar{\tau}}}{{\rm d} E_\gamma}$ 
are completely determined once the dark matter mass is given.
Here we simulate the gamma-ray spectrum of each final state through Pythia 8~\cite{Sjostrand:2007gs}, and verify that our results agree with the PPPC4DMID~\cite{Cirelli:2010xx}. 
The shape of the gamma-ray spectrum are the combination of the spectra of each final state, governed by the ratio $R$. 
Once the relative weight $R$ is fixed, the shape of the gamma-ray flux at given dark matter mass is thus determined. 
To determine the total normalization of the flux
\bea
        \frac{1}{8\pi m_X^2}  \langle \sigma v\rangle 
       	 \int_{\Delta \Omega} J(\psi) {\rm d}\Omega,
\eea
we need to know the dark matter mass $m_X$, the total thermal cross section $\sigma v$ and the integration over the $J$ factor.
The integration over the the  $J$ factor is a constant
since the gamma-ray flux data in Ref.~\cite{Daylan:2014rsa} are normalized at an angle of $5^{\circ}$ from the Galactic Center with the generalized NFW halo profile with an inner slope of $\gamma = 1.26$. 
Therefore, to accommodate the observed spectrum of the gamma-ray excess,  we need to determine the dark matter mass $m_X$ and thermally averaged cross section $\sigma v$ at given ratio $R$.

\begin{figure}[htp]
\begin{center}
\includegraphics[width=0.22\textwidth]{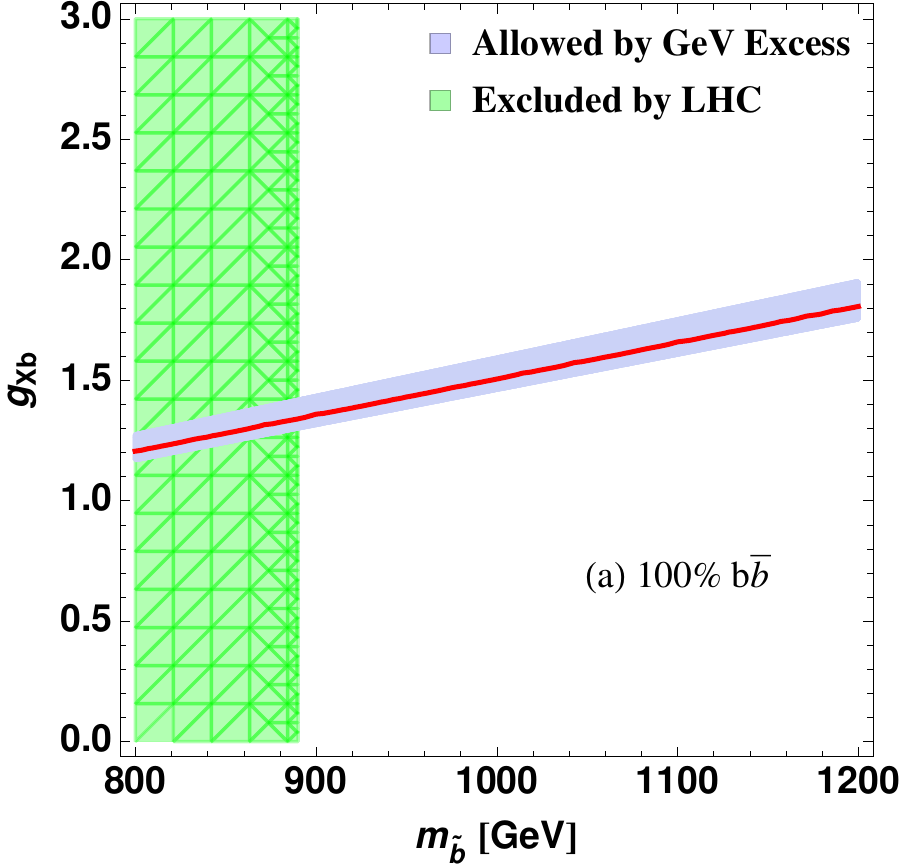} 
\includegraphics[width=0.22\textwidth]{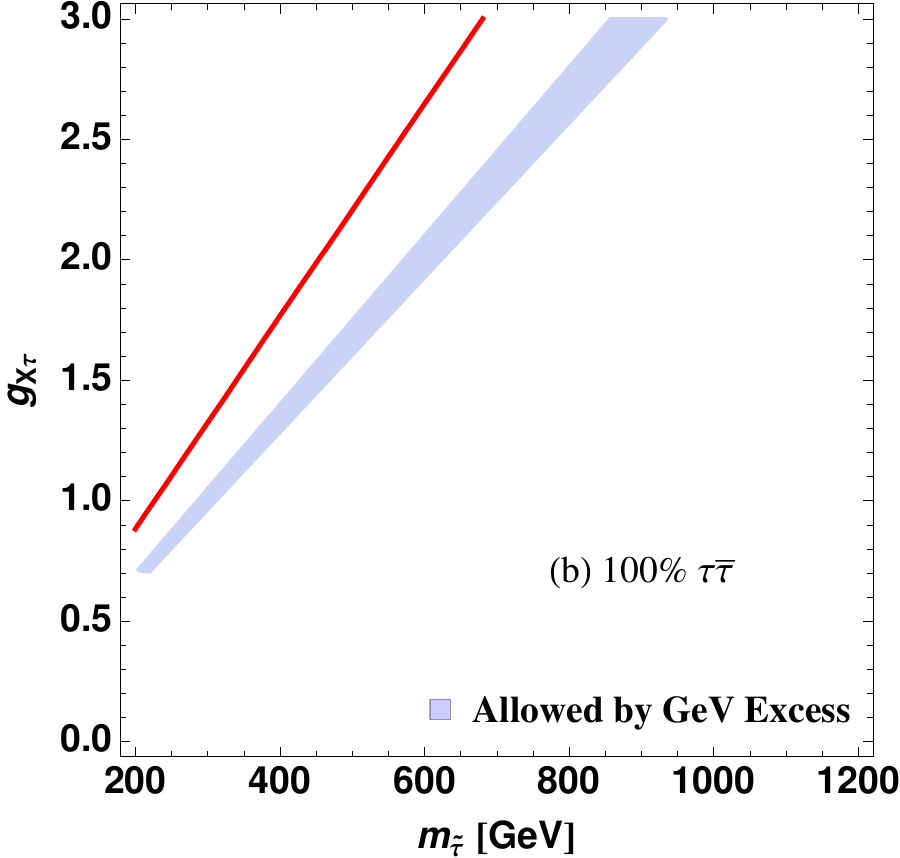} 
\caption{\small (a) The regions of the $(m_X, g_{Xb})$ plane in the quark portal model, for the $35.9$ GeV dark matter which gives the best fit of the quark portal dark matter. Both the regions (light blue) allowed by the gamma-ray excess and the contour (red line) which gives the correct relic density are shown. We also display the region excluded by the direct searches at the LHC. (b) The regions of the $(m_X, g_{X\tau})$ plane in the lepton portal model, for the $6.9$ GeV dark matter which gives the best fit of the lepton portal dark matter. Both the regions (light blue) allowed by the gamma-ray excess and the contour (red line) which gives the correct relic density are shown.  }
\label{fig:excess1}
\end{center}
\end{figure}

Given the relative weight $R$, we fit the resulting gamma-ray spectrum to the extract data shown in Ref.~\cite{Daylan:2014rsa}.
To perform this global fit, we define a $\chi^2$ statistic as
\bea	
	\chi^2 = \sum_i \frac{({\mathcal O}_i - \Phi_i^\gamma)^2}{\sigma_i^2},
\eea
where ${\mathcal O}$ and $\sigma$ are the extracted data and errors from Fig. 5 in Ref.~\cite{Daylan:2014rsa},
and $\Phi^\gamma$ is the predicted gamma-ray flux.
Then the global $\chi^2$ fit is performed to extract out the best values of the parameters ($m_X, \sigma v$).
In Fig.~\ref{fig:flux}, we show the best-fit spectra of the gamma-rays, for several values of the relative weight $R = 0, 0.25, 0.57, 0.8, 1$. 
From the Fig.~\ref{fig:flux}, we note that the larger the relative weight $R$ the better fit of the data.
On the left panel of Fig.~\ref{fig:chi}, we show contours of the dark matter mass and annihilation cross section required to fit of the gamma-ray spectrum at the 68\%, 90\%, and 95\% CLs, for a variety of the relative weight $R$.

To check whether the thermally averaged cross section is compatible to the relic density,
we list the allowed region for the thermally averaged cross section at the  95\% CL at given mass range $(6, 36)$ GeV of the DM. 
The procedure is as follows. First for each value of the relative weight $R$, we obtain the best fit values and the  95\% CL contour of the DM mass and the thermally averaged cross section. 
Then we project the 95\% CL contour to the obtain the allowed region of the thermally averaged cross section at the best value of the DM mass. 
Finally we scan over all the possible combinations of the $b\bar{b}$ and $\tau\bar{\tau}$ final states.
The results are shown on the right panel of the Fig.~\ref{fig:chi}.  
In Fig.~\ref{fig:chi}, it shows the allowed range of the thermally averaged cross section at given relative weight $R$, which corresponds to the best value of the DM mass mass$m_X$. 
To accommodate the observed spectrum of the gamma-ray excess, 
the DM annihilates in the low-velocity limit with a cross section of
\bea
	\langle \sigma v \rangle = (0.5 \sim 2.0) \times 10^{-26} \textrm{ cm}^3/s,
\eea
depending on the dark matter mass.
On the other hand, to satisfy the relic abundance, the DM cross section is limited to be in a very narrow range
\bea
	\langle \sigma v \rangle = (1.5 \sim 1.7) \times 10^{-26} \textrm{ cm}^3/s.
\eea
The overlap in the parameter space between the gamma-ray excess and the relic density is the favored region: the dark matter mass in the range of $(24, 36)$ GeV, and the corresponding ratio $R$ in the range of $(0.67, 1)$.
The lepton portal dark matter model can not explain the observed gamma-ray excess, while the quark portal dark matter model could accommodate the excess.
For the fermion portal dark matter, to explain the gamma-ray excess and satisfy the relic density,  
the $b\bar{b}$ final state should dominate over the $\tau\bar{\tau}$ final state. 

\begin{figure}[htp]
\begin{center}
\includegraphics[width=0.22\textwidth]{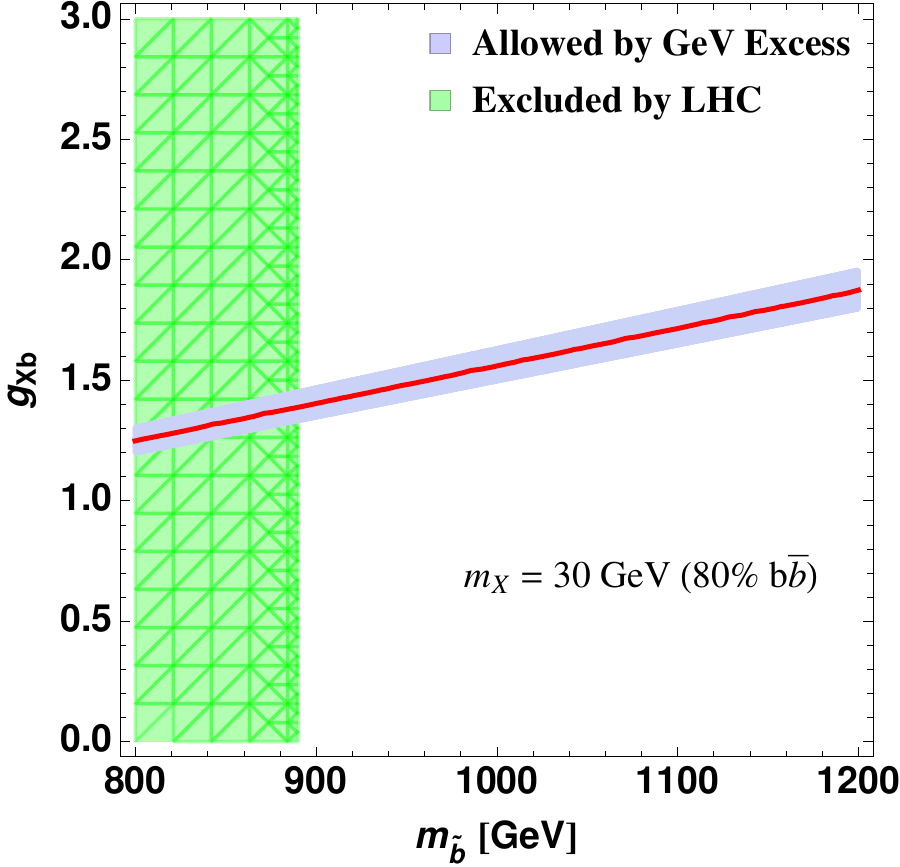} 
\includegraphics[width=0.22\textwidth]{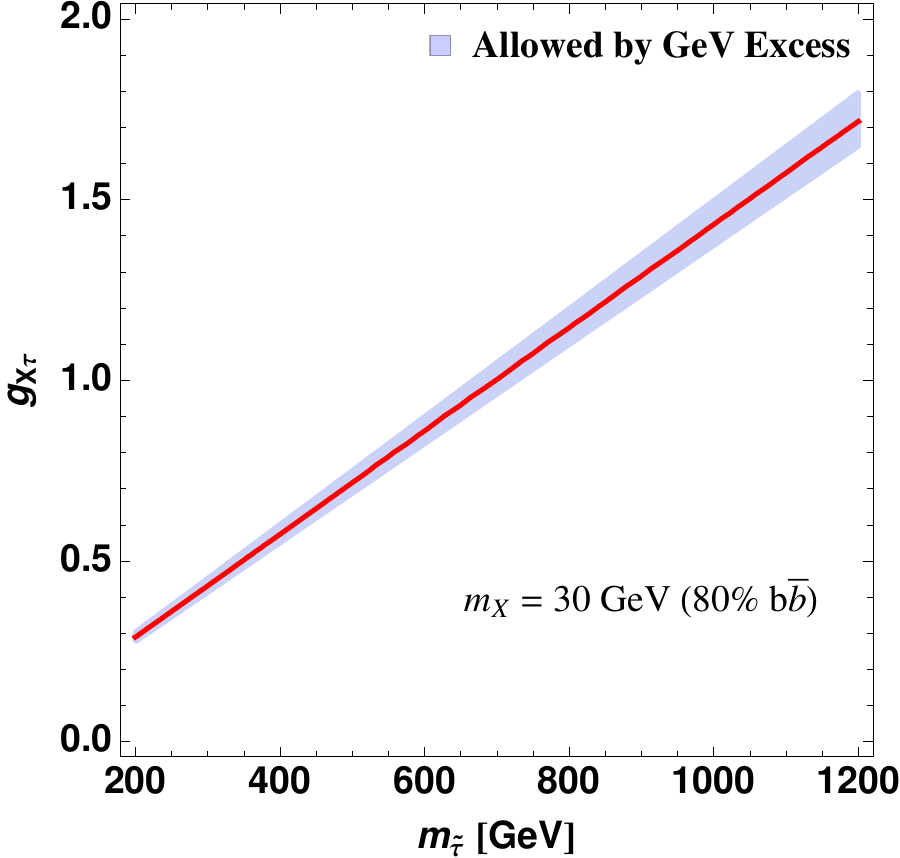}
\caption{\small The regions (light blue) allowed by the gamma-ray excess and the contour (red line) which gives the correct relic density in the fermion portal model. The vector dark matter mass is taken as 30 GeV, which corresponds to the the best fit from the combination of 80\% $b\bar{b}$ and 20\% $\tau\bar{\tau}$. Left: the $(m_X, g_{Xb})$ plane.  Right: the $(m_X, g_{X\tau})$ plane.  We also display the region excluded by the direct searches at the LHC on the left panel. }
\label{fig:excess2}
\end{center}
\end{figure}

\begin{figure}[htp]
\begin{center}
\includegraphics[width=0.22\textwidth]{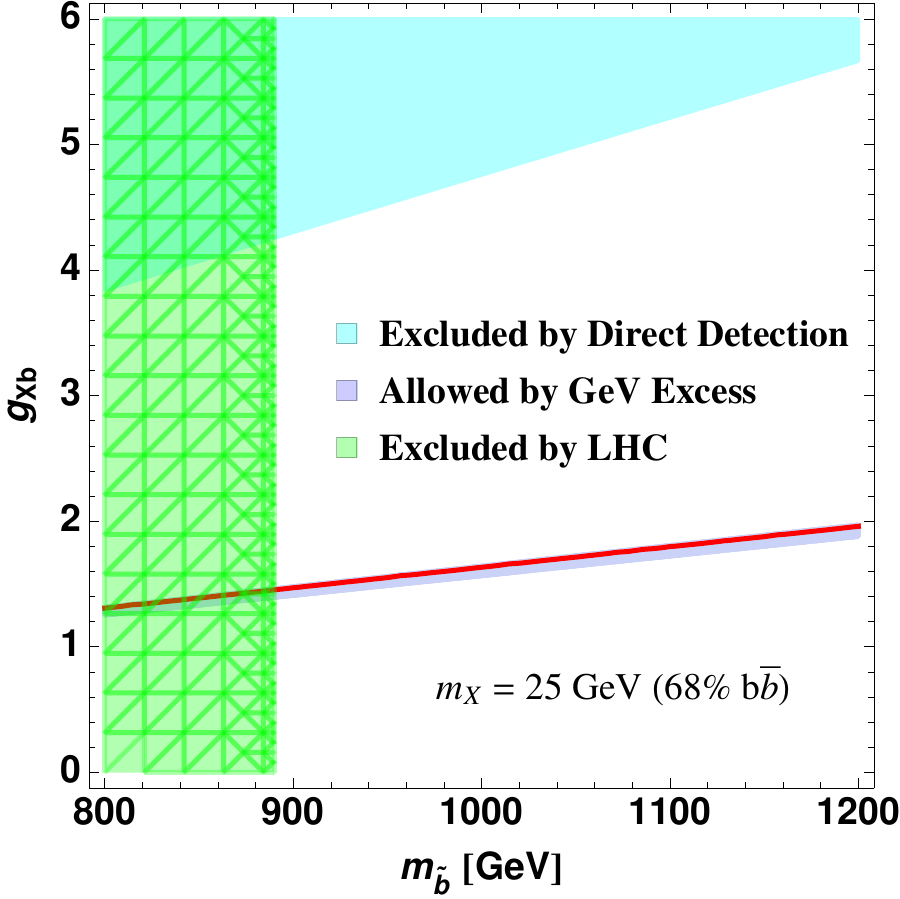} 
\includegraphics[width=0.22\textwidth]{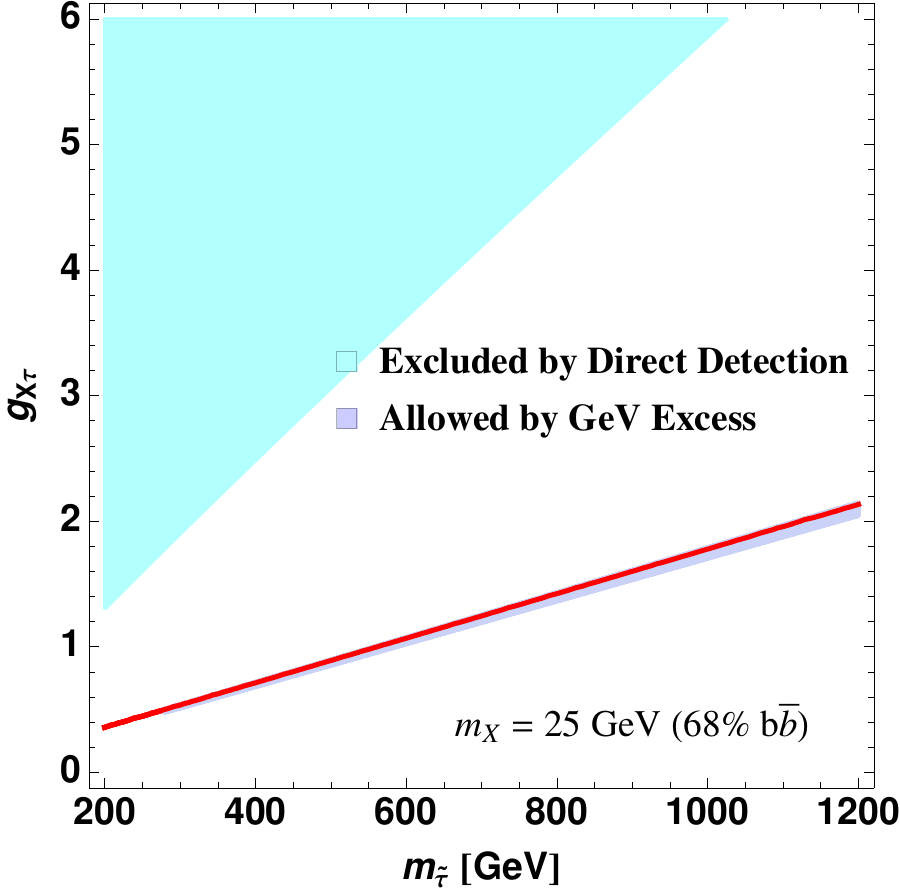} 
\caption{\small Similar to as shown in Fig.~\ref{fig:excess2}, but for a dark matter with mass $25$ GeV. We also show the excluded region from the direct detection experiments. }
\label{fig:excess3}
\end{center}
\end{figure}

We could covert the requirements on the thermally averaged cross section to the constraints on the model parameters. 
As we know, the $t$-channel process is dominant in the calculation of the relic density and the gamma ray flux.
Therefore, the requirements on the gamma-ray excess and the relic density put constraints on the fermion portal mediator masses and the couplings $g_X$.
We also place the constraints from the direct detection experiments, and the LHC search limits.
In the quark flavor model, the dark matter annihilates into  100\%  $b\bar{b}$ final state. 
The values of the best fit in this model is $m_X = 35.9 $ GeV, and $\sigma v = 1.73 \times 10^{-26} \textrm{ cm}^3/s$. 
Fig.~\ref{fig:excess1} (a) shows the parameter space allowed by the gamma-ray excess at the 95\% CL for a $35.9 $ GeV DM. 
In this model, the relic density is completely compatible to the allowed parameter region.
However, pure $b\bar{b}$ final state is  in tension with the constraints from the antiproton flux in the indirect detection~\cite{Cirelli:2014lwa}.
On the other hand, in the lepton portal model, as shown in fig.~\ref{fig:excess1} (b), there is no overlap region in the parameter space to satisfy the relic density and the explanation of the gamma-ray excess.  
Therefore, the lepton flavor model cannot explain the gamma-ray excess.
In the fermion portal model, %
we know that if the relative weight $R$ is greater than $68$\%, the relic density is compatible with the parameter space allowed by the gamma-ray excess. 
we show two benchmarks with the relative weight $R = 0.8 $ and $0.68$ in Figs.~\ref{fig:excess2} and~\ref{fig:excess3}. 
For the DM annihilating into the combination of 80\% $b\bar{b}$ and 20\% $\tau\bar{\tau}$ final states, we perform the global fit, and find that the values of the best fit are $m_X = 30 $ GeV, and $\sigma v = 1.63 \times 10^{-26} \textrm{ cm}^3/s$. 
Converting the allowed regions at the 95\% CL of the thermally averaged cross section for a 30 GeV DM back to the parameters of the model, we obtain the allowed parameter space in the $(m_{b'}, g_{Xb})$ plane and the $(m_{\tau'}, g_{X\tau})$ plane  shown in Fig.~\ref{fig:excess2}.
We also show the contour which gives the correct relic density in Fig.~\ref{fig:excess2}.
Similarly, we also show the allowed parameter space for a 25 GeV dark matter in Fig.~\ref{fig:excess3} with combined constraints from the relic density, the direct detection experiments and the collider searches.
From Figs.~\ref{fig:excess2} and~\ref{fig:excess3}, we note that, to both explain the gamma-ray excess and achieve the correct relic density, 
the coupling $g_{Xb}$ can be as smaller as $1.2$, while the coupling $g_{X\tau}$ can be as smaller as $0.2$, depending on the mediator masses.
In the fermion flavored dark matter, the coupling of the fermion dark matter to the bottom quark needs to be greater than $2$ to both explain the gamma-ray excess and satisfy the relic density.
The reason is that the $t$-channel annihilation cross section of the vector dark matter is larger than the one in the fermion dark matter case, at given  coupling strength and DM and its mediator masses.
Therefore, it is more attactive to explain the gamma-ray excess in the vector dark matter model.


\section{Conclusion}

In this study, we presented a fermion portal dark matter model, with a spin-$1$ vector dark matter and fermion 
mediators. 
In this model, the vector dark matter couples to the SM fermions through the fermion portal.
We assume that the vector dark matter predominately couples to the third generation fermions to explain the observed gamma-ray excess, and to evade the direct detection constraints.
We performed a detail relic density calculation, and showed that the $t$-channel process $XX \to f\bar{f}$ 
dominates the relic density for a light dark matter.
Constraints from the direct detection experiments, electroweak precision tests, the Higgs invisible decay, and collider searches were also investigated. 
We found that the regions of the parameters in the scalar sector are tightly constrained by the spin-independent direct detection experiments, and the Higgs invisible decay.
On the other hand, the constraints on the parameters in the fermion portal sector are quite weak, because the fermion portal sector only contributes to the 
spin-dependent cross section through the triangle loop diagrams. 
At the LHC, the direct searches on the final states with the bottom (top) pair plus missing transverse energy 
place very strong constraints on the quark mediator masses.  
To explain the observed gamma-ray excess, 
we studied how the spectra vary with respect to the combinations of the $b\bar{b}$ and $\tau\bar{\tau}$ final states, and determined which combinations of the dark matter mass and final states provide the best fit to the observed gamma-ray excess. 
Our results indicated that the pure $\tau\bar{\tau}$ final state cannot be consistent with the relic density, and the pure $b\bar{b}$ final state is in tension with constraints from the antiproton flux in the indirect detection.
However, the combined $b\bar{b}$ and $\tau\bar{\tau}$ final states with $R > 0.67$ are consistent with all the constraints. 
We concluded that still a large range of the parameter space could provide excellent explanation of the gamma-ray excess in galactic center and achieve the correct relic density at the same time.
Although the dark matter origin of the observed gamma-ray excess needs additional support, 
the combination of the indirect, direct and collider searches in future may unveil the underlying interactions of the vector dark matter, and identify the possible signatures of the vector dark matter.

\section*{Acknowledgements}

We would like to thank Wei Xue, Can Kilic, Jacques Distler, and Duane Dicus for very helpful discussions and valuable comments on the manuscript. 
The research was supported by the National Science Foundation under Grant Numbers PHY-1315983 and PHY-1316033.


\appendix


\begin{widetext}

\section{A Vector Fermion-Portal Model}

In this appendix, we present a vector fermion-portal model in detail. 
The gauge group is $SU(2)_L \times U(1)_1 \times U(1)_2$ with the gauge coupling $g_1 = g_2 = g'$. 
The Higgs doublet $H$ with the electroweak vacuum expectation value (VEV) $v$ and a complex scalar $\phi$ with a VEV $u$ are introduced to break the symmetry down to the $U(1)_{\rm em}$ group.
The breaking pattern is determined from the scale of the VEV $u$. If the scale $u > v$,
the gauge symmetry breaking is, first $U(1)_1 \times U(1)_2 \to U(1)_Y$ and then $ SU(2)_L \times U(1)_Y \to U(1)_{\rm em}$.
On the other hand, if the scale $v > u$, 
first $SU(2)_L \times U(1)_{1 \oplus 2} \to U(1)_{\rm em}$ where the diagonal subgroup $U(1)_{1 \oplus 2}$ of the $U(1)_1 \times U(1)_2$ is identified as the  hypercharge group $U(1)_Y$, and then off-diagonal subgroup $U(1)_{1\ominus 2}$ breaks to nothing due to the complex scalar gets the VEV $u$.
%
%

The scalar sector of the model is
\bea
	{\mathcal L}  = D_\mu H^\dagger D^\mu H + D_\mu \phi^* D^\mu \phi - V(H, \phi),
\eea
where the covariant derivatives are
\bea
	D^\mu H     &=& \partial^\mu H + i g_2\frac{\tau^a}{2} W^{a\mu} H + i \frac{g'}{2} (B_1^\mu + B_2^\mu ) H,\\
	D^\mu \phi  &=&  \partial^\mu \phi + i \frac{g'}{2} (B_1^\mu - B_2^\mu ) \phi.
\eea
The general scalar potential can be written as
\bea
	V(H, \phi) &=&  \lambda_H (H^\dagger H - v^2/2)^2 + \lambda_S (\phi^* \phi - u^2/2)^2 \nn\\
	&&   + \lambda_{SH} (\phi^* \phi -u^2/2) ( H^\dagger H - v^2/2).
\eea
Similar to the $T$-parity~\cite{Cheng:2003ju,Cheng:2004yc} in the Little Higgs model, a parity is assigned to have $B_1^\mu \leftrightarrow B_2^\mu$. 
Let us define the following combination 
\bea
	X^\mu &=&\frac{1}{\sqrt{2}}(B_1^\mu - B_2^\mu),\\
	B^\mu &=&\frac{1}{\sqrt{2}}(B_1^\mu + B_2^\mu),
\eea
where we identify  $B^\mu$ as the $U(1)_Y$ gauge field in the SM.
Under this parity, there are $X_\mu \to -X_\mu$. 
In this hidden Higgs mechanism, under the parity $\phi \to \phi^*$ is required to make sure its Goldstone component is absorbed after the spontaneous symmetry breaking.
Similar to SM, the dangerous $\partial^\mu \phi^* \phi X_\mu$ terms which make the $X_\mu$ instable,
disappear after the absorption of the Goldstone boson $\textrm{Im}\,\phi$ by the $X_\mu$. 
Therefore, all the couplings involved in the $X_\mu$ field is in pairs, and thus  $X_\mu$ is a stable dark matter candidate.


%
After symmetry breaking, there are mass mixing between the neutral component of the doublet ${\textrm Re}\,H_0$ and the real component of the scalar ${\textrm Re}\,\phi$.  
The mixing matrix is
\bea
	{\mathcal M}^2_{S} = \left( \begin{array}{cc}
						2 \lambda_H v^2 & \lambda_{SH} v u\\
						\lambda_{SH} v u &2 \lambda_S u^2 
						\end{array}\right).
\eea
Diagonalizing the above matrix, we obtain the mass squared eigenvalues
\bea
	m^2_{h,S} &=& \lambda_H v^2 + \lambda_S u^2 \mp 
		  \sqrt{(\lambda_S u^2 - \lambda_H v^2)^2 +  \lambda_{SH}^2 u^2 v^2 }.\nn\\
		  \label{eq:scalarmass}
\eea
and the eigenvectors $(h, S)$
\bea
	\left( \begin{array}{c} h \\ S \end{array}\right) = 
	\left( \begin{array}{cc}
	\cos\varphi & -\sin\varphi\\
	\sin\varphi & \cos\varphi
	\end{array}\right)
	\left( \begin{array}{c}  {\textrm Re}\,H_0 \\ {\textrm Re}\,\phi \end{array}\right), 
\eea
where the mixing angle  $\varphi$ is given by
\bea
	\tan 2\varphi = \frac{\lambda_{SH} u v}{\lambda_S u^2 - \lambda_H v^2}.
	\label{eq:scalarangle}
\eea

In the fermion sector, two right-handed fermion singlets $\psi_1$ and $\psi_2$ are introduced. The quantum numbers are  $(1, Y_1, Y_2)$ and $(1, Y_2, Y_1)$ respectively. The Lagrangian is written as
\bea
\mathcal{L} = \overline{\psi}_1 i\gamma_\mu D^\mu {\psi}_1 + \overline{\psi}_2 i\gamma^\mu D^\mu {\psi}_2,
\eea
where the covariant derivatives are
\bea
	D^\mu {\psi}_1 &=& \left[\partial^\mu + i g^\prime (Y_1 B_1^\mu +Y_2 B_2^\mu  )\right] {\psi}_1,\\
	D^\mu {\psi}_2 &=& \left[\partial_\mu + ig^\prime (Y_2 B_{1}^\mu +Y_1 B_2^\mu )\right] {\psi}_2.
\eea
The parity also has $\psi_1 \leftrightarrow - \psi_2$. 
Under this symmetry, the above Lagrangian is invariant.
So we introduce the combination
\bea
	F_R   &=&\frac{1}{\sqrt{2}}(\psi_1 + \psi_2),\\
	f_R   &=&\frac{1}{\sqrt{2}}(\psi_1 - \psi_2).
\eea
If we assign the quantum number $\frac{Y_1 + Y_2}{2}$ as the SM hypercharge $Y$, we could identify $f_R$ as the SM right-handed fermion.
Under this parity, we have $F_R  \to -F_R$, and $f_R \to f_R$.
According to the new combination, 
we rewrite the kinetic term as
\bea
	\mathcal{L} &=& \overline{f}_R i\gamma_\mu \left[\partial^\mu + i g^\prime Y B^\mu \right]  {f}_R   +\overline{F}_R i\gamma_\mu \left[\partial^\mu + i g^\prime Y B^\mu \right]  {F}_R \nn\\
	&& + \overline{F}_R i\gamma_\mu \left[ i g^\prime Y^\prime X^\mu \right]  {f}_R   + \overline{f}_R i\gamma_\mu \left[i g^\prime Y^\prime X^\mu \right]  {F}_R,
\eea
where $Y^\prime = \frac{Y_1 - Y_2}{2}$ is defined. 
%
%
In our setup, we introduce a left-handed singlet $F_L$ which has $F_L \to - F_L $ under the parity symmetry. 
Hence, the left-handed singlet $F_L$ and the right-handed fermion $F_R$ is combined together to form a vectorlike fermion singlet: $F = (F_L, F_R)$, with a heavy Dirac mass $M$.  
Therefore, for each SM fermion $f$, there is a fermion partner $F$. 
The interaction between the $f$ and $F$ is through the new gauge boson $X$, where the coupling strength is $g'Y^\prime$. 
Since the SM fermion $f$ interacts with the vector boson dark matter through the mediator $F$, it is a kind of the  fermion portal dark matters.

\section{Thermally Averaged Cross Sections}
\label{sec:appen1}

The thermal averaged annihilation cross section times relative velocity $\langle \sigma v \rangle$ is~\cite{Edsjo:1997bg}
\bea
	\langle \sigma v \rangle &=& \frac{1}{n_{\rm EQ}^2} \frac{g_i^2 T}{64\pi^4}
	 \int^{\infty}_{4 m_{\tmbox{X}}^2 } \left(\sigma v\right) s \sqrt{s-4m_{\tmbox{X}}^2} K_1\left(\frac{\sqrt{s}}{T}\right) {\rm d} s,
\eea 
and the number density at thermal equilibrium $n^2_{\rm EQ}$  is
\bea
	n_{\rm EQ} = \frac{g_i T}{2\pi^2}  m_{\tmbox{X}}^2 K_2(x),
\eea
where $g_i$ is internal degrees of freedom, the variable $x \equiv \frac{m_{\tmbox{X}} }{T}$,
and  the functions $K_1$ and $K_2$ are the modified Bessel function of the first kind and second kind.

The relevant couplings are given by
\bea
g_{XXh}&= - i g_\phi m_X s_\varphi, \qquad g_{XXS}= i g_\phi m_X c_\varphi,\\
g_{WWh}& =  i g m_W c_\varphi, \qquad g_{WWS} = i g m_W s_\varphi,\\
g_{ZZh}& =  i \frac{g}{c_W} m_Z c_\varphi , \qquad g_{ZZS} = i \frac{g}{c_W} m_Z s_\varphi.
\eea
 
The velocity times DM annihilation cross sections into $ff$ boson pairs before thermal average are given by 
\bea
	\left( \sigma v_{\rm rel} \right)_{t\textrm{-ch}} &=& \frac{N_c g_X^4}{72 \pi  s}\left\{ 
			    \frac{ 4 R^2 }{ R + \gamma (R-1)^2 } - (3 R^2+8 ) 
				\right.\nn\\			 
	  && \left. + 2\tanh^{-1}\left[\frac{\beta}{1+2(R-1) \gamma }\right] 
	 \frac{ R^2 + 4 - 12 R^2 \gamma + 2\gamma (3R^2 + 8) \left[R  + \gamma (R-1)^2 \right]   }{\beta (1+2(R-1)\gamma)}
	   \right\},
\eea
where the ratio $R = m_F^2/m_X^2$, $\beta = \sqrt{ 1 - 4 m_X^2/s}$, and $\gamma = m_X^2/s$.
The $s$-channel is
\bea \label{sv-ff}
\left( \sigma v_{\rm rel} \right)_{f\bar{f}}
=  \frac{N_c m_f^2}{36\pi v^2}\left| \sum_i \frac{g_{XXi}^2 }{s-m_i^2 + 
i m_i \Gamma_i} \right|^2  \left( 3 + \frac{s(s - 4 m_X^2)}{4 m_X^4} \right) \left( 1-\frac{4 m_f^2}{s} \right)^{3/2}   ,
\eea
where  
the involved couplings are $g_{XXi}$ with the index $i = h, S$.
There is no interference term since the fermion portal coupling is pure chirally right-handed.
\bea 
(\sigma v)_{\textrm{t-channel}} \simeq  \frac{2 N_c g_X^4}{9 \pi } \frac{m_X^2}{(m_{X}^2 + m_F^2)^2} + v^2 \frac{N_c g_X^4}{54 \pi } \frac{m_X^2 (7 m_X^4 + 10 m_F^2 m_X^2 - 5 m_F^4)}{(m_{X}^2 + m_F^2)^4} .
\eea

The velocity times DM annihilation cross sections into $W,Z$ boson pairs before thermal average are given by
\bea
\left( \sigma v_{\rm rel} \right)_{VV} = \frac{\delta_{VZ}}{72 \pi s} 
\left| \sum_i \frac{g_{XXi}^2g_{VVi}^2}{s-m_i^2 + 
i m_i \Gamma_i} \right|^2    \left( 3 + \frac{s(s - 4 m_V^2)}{4 m_V^4} \right) \left( 3 + \frac{s(s - 4 m_X^2)}{4 m_X^4} \right) \sqrt{1-\frac{4 m_V^2}{s}},
\eea
where $\delta_{VZ} = 1/2$ for the $Z$ boson, and  $\delta_{VZ} = 1$ for the $W$ boson.



\section{Effective Operators of the WIMP-nucleon Scattering}

For a real vector dark matter, the following contact operators for interaction with quarks and gluons are possible~\cite{Freytsis:2010ne,Yu:2011by}:
\bea
	{\mathcal O}_1^q  &=& X_\mu X^\mu \overline{q} q, \\
	{\mathcal O}_2^q  &=& X_\mu X^\mu \overline{q} \gamma^5 q, \\
	{\mathcal O}_3^q  &=& X_\nu \partial^\nu X_\mu \overline{q} \gamma^\mu q, \\
	{\mathcal O}_4^q  &=& X_\nu \partial^\nu X_\mu \overline{q} \gamma^\mu \gamma^5 q, \\
	{\mathcal O}_5^q  &=& \epsilon^{\mu\nu\rho\sigma}X_\nu \partial_\rho X_\sigma \overline{q} \gamma_\mu q, \\
	{\mathcal O}_6^q  &=& \epsilon^{\mu\nu\rho\sigma}X_\nu \partial_\rho X_\sigma \overline{q} \gamma_\mu\gamma^5 q, \\
	{\mathcal O}_7^q  &=& \frac{\alpha_s}{12\pi} X_\mu X^\mu  G^a_{\mu\nu}  G^a_{\mu\nu}, \\
	{\mathcal O}_8^q  &=& \frac{\alpha_s}{8\pi} X_\mu X^\mu  G^a_{\mu\nu}  \tilde{G}^a_{\mu\nu}.
\eea
The effective Lagrangian at the quark-gluon level is
\bea
	{\mathcal L}_{\rm eff} = \sum_{k=1}^8 c^q_k {\mathcal O}_i^q,
\eea
where $c^q_i$ are coefficients of the operators.
These operators induce effective Lagrangian at the nucleon level
\bea
	{\mathcal L}_{\rm eff} = \sum_{k=1}^6 c^N_k {\mathcal O}_i^N,
\eea
where the ${\mathcal O}_i^N$ with $N= p,n$ are
\bea
	{\mathcal O}_1^N  &=& X_\mu X^\mu \overline{N} N, \\
	{\mathcal O}_2^N  &=& X_\mu X^\mu \overline{N} \gamma^5 N, \\
	{\mathcal O}_3^N  &=& X_\nu \partial^\nu X_\mu \overline{N} \gamma^\mu N, \\
	{\mathcal O}_4^N  &=& X_\nu \partial^\nu X_\mu \overline{N} \gamma^\mu \gamma^5 N, \\
	{\mathcal O}_5^N  &=& \epsilon^{\mu\nu\rho\sigma}X_\nu \partial_\rho X_\sigma \overline{N} \gamma_\mu N, \\
	{\mathcal O}_6^N  &=& \epsilon^{\mu\nu\rho\sigma}X_\nu \partial_\rho X_\sigma \overline{N} \gamma_\mu\gamma^5 N,
	\eea
The above operators has contributions to both SI and SD cross sections with suppression factors $v^2$ and/or $q^2$~\cite{Kumar:2013iva}, except ${\mathcal O}_1^N$ and ${\mathcal O}_6^N$.
Therefore, we will only consider the two relevant operators in the following.
In the non-relativistic limit, the leading contributions to the operators ${\mathcal O}_1^N$ and ${\mathcal O}_6^N$ are
\bea
	{\mathcal O}_1^N &\simeq& 2 m_N ,\\
	{\mathcal O}_6^N &\simeq& 8 m_X m_N \vec{s}_X\cdot \vec{s}_N,
\eea
where $\vec{s}_N$ is the nucleon spin and $\vec{s}_X$ the DM spin.
Therefore the operator ${\mathcal O}_1^N$ has the leading contribution to the SI cross section, while
the operator ${\mathcal O}_6^N$ has the leading contribution to the SD cross section.

There are connections between the coefficients $c^q$ at the quark and gluon level and the coefficients $c^N$ at the nucleon level~\cite{DelNobile:2013sia}. 
For the vector dark matter, the connections between $c^N_{1,6}$ and $c^q_{1,6,7}$ are 
\bea
	c^N_{1} &=& \sum_{q = u, d, s} c^q_{1} \frac{m_N}{m_q} f_{Tq}^{(N)} + \frac{2}{27} f_{TG}^{(N)} \left( \sum_{q = c, b, t} c^q_{1, 2} \frac{m_N}{m_q} - c^q_{7} m_N \right), \\
	c^N_{6} &=& \sum_q c^q_{6} \, \Delta_q^{(N)}.
\eea 
Here the quantity $f_{Tq}^{(N)}$ is defined by the matrix element of the light quark ($q$) bilinear with the nucleon $N$: 
\bea
	 f_{Tq}^{(N)}  \equiv \langle N| \frac{m_q}{m_N}\overline{q} q |N \rangle ,
\eea
and $f_{TG}^{(N)}$ similarly by the gluon operators:
\bea
	f_{TG}^{(N)}  \equiv  - \frac{27}{2m_N} \langle N|  \frac{\alpha_s}{12\pi} G^{a\mu\nu}G^a_{\mu\nu} |N \rangle.
\eea
The gluon contribution can be expressed in terms of light quarks via
\bea
	f_{TG}^{(N)} = 1 - \sum_{q=u,d,s} f_{Tq}^{(N)}.
\eea
We adopt the following values to describe the nuclear quark content: $f^p_{Tu}  = 0.015$, $f^p_{Td}  = 0.019$ for the proton and $f^n_{Tu} = 0.011$, $f^n_{Td} = 0.027$ for the neutron,  and $f^p_{Ts} = f^n_{Ts} = 0.045$~\cite{Belanger:2013oya}. Therefore, $f_{TG} = 0.92$ implies that heavy quark contribution dominates over the light quarks.
The quantity $\Delta^N_q$ is determined by the matrix element of the axial vector current with the nucleon $N$ is 
\bea
	 \Delta^N_q \equiv \langle N |\bar{q} \gamma^\mu \gamma^5 q | N\rangle s^\mu. 
\eea
where $s^\mu $ is the nucleon spin four-vector.
We take the following values to describe the nuclear quark content: $\Delta^p_u = \Delta^n_d = 0.843$, $\Delta^p_d = \Delta^n_u = 0.427$, and $\Delta^p_s = \Delta^n_s = -0.085$~\cite{Belanger:2013oya}. This result is consistent with the measurement of the axial-vector current form factor~\cite{Schindler:2006jq}, encoded the nucleon spin structure.

\section{Calculations on Loop-induced Vertices}

%

\subsection{Triangle Loop Diagrams}

The general Lorentz structure of the vertex~\cite{Gounaris:2000tb} is
\bea
	\Gamma^{\alpha \beta \mu} (p_1, p_2) = \frac{iq^2}{m_X^2}
	\left[ f_1 (p^\alpha g^{\beta\mu}+ p^\beta g^{\alpha\mu}) - f_2 \epsilon^{\mu\alpha\beta\rho }q_\rho\right],
\eea
where $p = p_1 + p_2$ and $q = p_1 - p_2$. Note that $f_1$ is the CP-violating operator, while $f_2$ is CP-conserving. 
The Fermion loop shown in Fig.~\ref{fig:vvvtri} contributes to the CP-conserving $f_2$ operator. In term of the Feynman integral, the operator is
\bea
	f_2 = N_c \frac{eQ g_X^2}{\pi^2}\int^1_0 {\rm d} x \int^{1-x}_0 {\rm d} y \frac{x y }{m_f^2 + (m_F^2 - m_f^2)(x+y) - q^2 x y + m_X^2 (x+y)(x+y-1)} + (m_f \leftrightarrow m_F),
\eea
In the limit of small $q^2$, we obtain 
\bea
	f_2 = N_c \frac{eQ g_X^2}{\pi^2} \frac16 \int_0^1  {\rm d} z \frac{z^3 }{m_f^2 + (m_F^2 - m_f^2)z + m_X^2 z(z-1)} + (m_f \leftrightarrow m_F).
\eea
In the zero external mass limit, we have 
\bea
	f_2 = N_c \frac{eQ g_X^2 }{\pi^2}\frac{  4 (m_F^6 - m_f^6) \log\frac{m_F}{m_f} - 3(m_F^2 + m_f^2) (m_F^2 - m_f^2)^2  }{12 (m_F^2 - m_f^2)^4}.
\eea
%

\subsection{Box Loop Diagrams}

The vector dark matter does not couple with the gluons at the tree level.
However, it does couple to the gluon through box diagrams at the loop level, as shown in Fig.~\ref{fig:vvvvbox}.  
The loop-induced effective Lagrangian is
\bea
	{\mathcal L} = b_g
	 B^{\rho}B_{\rho}G^{a\mu\nu}G^a_{\mu\nu}.
\eea
There are three diagrams contribute to the effective coupling strength $b_g$.
Here we seperate the three contributions of the diagram, (a), (b), and (c) into
\begin{equation}
b_g =
\frac{\alpha_s g_X^2}{8\pi} (f_a + f_b + f_c),
\end{equation}
where $f_{a,b,c}$ corresponds to the result of each diagram respectively. 
Here we present the results of each diagram~\cite{Hisano:2010yh} in below.
The first diagram (a) has
\bea
f_a & = &\frac{m_f^2}{12 m_X^4}\left( (m_f^2 +m_F^2 - m_X^2) L - \log\frac{m_f^2}{m_F^2}\right) \nn\\
&&+ \frac{1}{6 \Delta m_X^4} \left\{ m_X^4(m_F^2 - m_f^2) + m_f^2 m_X^2(5m_F^2 + m_f^2) + m_f^2 m_F^2 L \left[ 5 m_F^4 + 20 m_f^2 m_F^2 - m_f^4 +   m_X^2 (9m_F^2 + m_f^2  \right] \right\} \nn\\
&& + \frac{1}{ \Delta^2 m_X^4} \left\{ m_F^2 m_f^2 \left[ m_X^2(3m_F^2 + m_f^2) -  (m_F^2 -m_f^2)^2 \right]  \right.\nn\\
&&\left.
+ m_f^2m_F^4 L \left[ m_X^2(m_F^4+10m_f^2m_F^2+5m_f^4)-(m_F^2-m_f^2)^2(m_F^2+3m_f^2) \right] \right\},
\eea
where in the case $m_F > m_X $, one has
\bea
 \Delta &=& m_X^4-2m_X^2(m_f^2+m_F^2)+(m_F^2-m_f^2)^2, \\
 L &= &
	   \frac{1}{\sqrt{\Delta}}\ln
\biggl(\frac{m_F^2+m_f^2- m_X^2+\sqrt{\Delta}}
{m_F^2+m_f^2- m_X^2-\sqrt{\Delta}}\biggr).
\eea
The second diagram (b) has
\bea
f_b = f_a (m_f \to m_F).
\eea
The third diagram (c) has
\bea
f_c &=& \frac{1}{12 m_X^4} \left( (m_X^2 -m_f^2 - m_F^2)(m_f^2 +m_F^2) L + (m_f^2 - m_F^2) \log\frac{m_f^2}{m_F^2}\right)\nn\\
&& + \frac{1}{6 \Delta m_X^4} \left( - 2 m_X^6 + (m_f^2-m_F^2)^2 (2 m_f^2 m_F^2 L - m_X^2) + (m_f^2 +m_F^2)m_X^2 (3 m_X^2 - 4 m_f^2 m_F^2 L)  \right),	
\eea
which is symmetry under $m_f \to m_F$.
Since we consider the light dark matter, 
in the limit of zero external masses, we have
\bea
	f_G = \frac{m_F^4 - m_f^4 + 2 m_f^2 m_F^2 \log\frac{m_f^2}{m_F^2} } {2 (m_F^2 - m_f^2)^3}.
\eea
Taking another limit $m_f \to 0$, we obtain
\bea
	f_G = \frac{\alpha_s g_X^2}{8\pi}\frac{3m_F^2- 2 m_X^2}{6(m_F^2-m_X^2)^2}.
\eea

\end{widetext}



\end{document}